\documentclass[12pt]{article}           
\usepackage{graphicx, float, amssymb,latexsym}
\def\preprint{TUW-15-14}       
\def\title{Cosmic Acceleration as an Optical Illusion}

\long\def\abstract{ 
We consider light propagation in an inhomogeneous irrotational dust universe 
with vanishing cosmological constant, with initial conditions as in
standard linear perturbation theory.
A non-perturbative approach to the dynamics of such a universe is
combined with a distance formula based on the Sachs optical equations.
Then a numerical study implies a redshift-distance relation that
roughly agrees with observations.
Interpreted in the standard homogeneous setup, our results would
appear to imply the currently accepted values for the Hubble rate and the 
deceleration parameter;
furthermore there is consistency with density perturbations at last 
scattering.
The determination of these three quantities relies only on a single parameter
related to a cutoff scale.
Discrepancies with the existing literature are
related to subtleties of higher order perturbation theory which make both the
reliability of the present approach and the magnitude of perturbative effects
beyond second order hard to assess.
}

\def\topt{{\theta_\mathrm{opt}}}
\def\sopt{{\sigma_\mathrm{opt}}}
\def\rdot{\dot{\,}}

\def\Ht{H_\sharp}
\def\Hd{H_\mathrm{inf}}
\def\dS0{d_{S\sharp}}
\newcommand{\capt}[3]{{\caption{\label{#2} #3}}}
\newcommand{\fref}[1]{Fig.~\ref{#1}}

\catcode`\"=\active \let"=\"

\def\ifundefined#1{\expandafter\ifx\csname#1\endcsname\relax}
\def\bye{\end{document}}   
\long\def\new#1\endnew{{\bf #1}}
\long\def\del#1\enddel{} 

\def\HS#1 {\hspace*{#1pt}} \def\VS#1 {\vspace*{#1pt}}
\def\BC{\begin{center}}    
\def\EC{\end{center}}

\let\0=\over      \let\1=\vec      \def\2{{1\over2}}    \let\3=\ss
\let\4=\underline \let\5=\overline \let\6=\partial      \def\7#1{{#1}\llap{/}}
\def\8#1{{\textstyle{#1}}}         \def\9#1{{\ifmmode{\pmb{#1}}\else\bf#1\fi}}

          \def\({\left(}       \def\){\right)}

\def\eeql#1 {\label{#1}\eeq}      \let\nn=\nonumber  
\def\beq{\begin{equation}}      \def\eeq{\end{equation}}        
\def\bea{\begin{eqnarray}}      \def\eea{\end{eqnarray}}

\let\and=\wedge

\let\bra=\langle        \let\ket=\rangle        \def\<#1\>{\bra #1 \ket}

\def\rel#1 #2{\buildrel #1 \over {#2}}  
\def\fnote#1#2{\begingroup\def\thefootnote{#1}\footnote{#2}
                \addtocounter{footnote}{-1}\endgroup}   

\let\a=\alpha   \let\b=\beta    \let\g=\gamma   \let\d=\delta   %% --> GREEK
         \let\th=\theta  \let\e=\varepsilon
   \let\l=\lambda  \let\m=\mu      
\let\n=\nu                  \let\r=\rho     \let\s=\sigma 
                 
            \let\O=\Omega    
        \let\L=\Lambda  \let\G=\Gamma   \let\D=\Delta

  \def\cc{{\cal C}} \def\cd{{\cal D}}

  \def\co{{\cal O}}  
 \def\cR{{\cal R}}

\def\IR{{\mathbb R}}

\def\plb#1 #2 {Phys. Lett. {\bf B#1} #2 }
\def\phr#1 #2 {Phys. Rep. {\bf  #1} #2 }        
\def\npb#1 #2 {Nucl. Phys. {\bf B#1} #2 }
\def\aph#1 #2 {Ann. Phys. {\bf #1} #2 }         
\def\jmp#1 #2 {J. Math. Phys. {\bf #1} #2 }
\def\jgp#1 #2 {J. Geom. Phys. {\bf #1} #2 }
\def\prd#1 #2 {Phys. Rev. {\bf D#1} #2 }
\def\prl#1 #2 {Phys. Rev. Lett. {\bf #1} #2 }
\def\rmp#1 #2 {Rev. Mod. Phys.  {\bf #1} #2 }
\def\zpc#1 {Z. Phys. {\bf #1C} }
\def\cmp#1 #2 {Commun. Math. Phys. {\bf #1} #2 }
\def\cqg#1 #2 {Class.Quant.Grav. {\bf #1} #2 }
\def\mpl#1 {Mod. Phys. Lett. {\bf A#1} }
\def\cpc#1 {Computer Phys. Commun. {\bf #1} }   % Belfast,cpc@v1.am.qub.ac.uk
\def\ijmp#1 {Int. J. Mod. Phys. {\bf A#1} }
\def\ijmpC#1 {Int. J. Mod. Phys. {\bf C#1} }
\def\atmp#1 {Adv. Theor. Math. Phys. {\bf #1} }

\def\BP{\begin{picture}} \def\EP{\end{picture}}         %% --> PICTURE macros 
\newcounter{TRefNX} \let\OLDcite=\cite  \makeatletter%       DRAFT MODE macros
\def\makeTRefs#1{\@for  \NewTRef:=#1\do{\global\makeTRef{\NewTRef}}}
\def\makeTRef#1{\ifundefined{TRef#1}\stepcounter{TRefNX}%
\expandafter\xdef\csname TRef#1\endcsname{\theTRefNX}\fi}\makeatother
\def\NEWcite#1{\makeTRefs{#1}\OLDcite{#1}}  

\ifundefined{draftmode} {} 
\else        
   \let\cite=\NEWcite
   \newcount\HOUR  \HOUR=\the\time \divide\HOUR by 60   \multiply \HOUR by 60 
   \newcount\MIN   \MIN=\the\time  \advance\MIN by -\HOUR  \divide\HOUR by 60
   \ifnum  \MIN>9  \def\printTIME{{\it\the\HOUR\,:\,\the\MIN}}
   \else   \def\printTIME{{\it\the\HOUR\,:\,0\the\MIN}} 
   \fi % \printTIME
   
   \def\LLab#1{\BP(0,0)\unitlength=1mm\put(-12,.5){\makebox(0,0)[cr]{\small #1
        \rlap{$_{_{\makeatletter\csname TRef#1\endcsname\makeatother}}$}}}\EP}
\fi

\begin{document}

%\del

{\hfill\preprint }
\vskip 15mm
\begin{center} 
{\LARGE\bf   \title }\vskip 10mm
Harald Skarke\fnote{*}{e-mail: skarke@hep.itp.tuwien.ac.at}\\[3mm]
Institut f\"ur Theoretische Physik, Technische Universit\"at Wien\\
Wiedner Hauptstra\ss e 8--10, 1040 Wien, Austria
        
\vfill                  {\bf ABSTRACT } 
\end{center}    
\abstract

\vfill %\noindent \preprint\\[5pt] \finished \vspace*{9mm}
\thispagestyle{empty} \newpage
\pagestyle{plain}

\newpage
\setcounter{page}{1}
%\enddel
  
\section{Introduction}
The fact that cosmological observations do not conform to the predictions
of Friedmann-Lemaitre-Robertson-Walker (FLRW) models with a vanishing
cosmological constant $\L$ is usually interpreted as an indication that
$\L$ differs from zero.
Clearly our actual universe deviates from the idealized FLRW cases by
hosting inhomogeneities, and there have been many suggestions that
the latter might have effects which would explain the data %observations
without requiring $\L$; see e.g.~Ref.~\cite{astro-ph/0209584} for an early
proposal of this kind.
The main challenge for any such claim is to explain why we perceive 
an accelerated expansion.
Basically there are two possible routes as well as combinations of them.
On the one hand the inhomogeneities might have an impact on
the actual expansion of the universe (suitably defined in
terms of the evolution of volumes of spatial regions).
On the other hand there is the possibility that they %re are effects on
affect light propagation in a subtle way which modifies the usual 
distance-redshift relations.
In the present work we are mainly concerned with the second scenario, which 
relies on the obvious yet important insight that almost every single piece of 
evidence on the evolution of the cosmos relies on the observation of photons 
with telescopes or other devices; Ref.~\cite{1103.5331} provides a particularly 
forceful presentation of this point.

There is an extensive amount of literature on light propagation in the presence
of inhomogeneities; see 
e.g.~Refs.~\cite{0711.4264,0812.2872,1109.2484,1203.4479,1204.0909,%
1207.2109,1401.7973,Umeh:2014ana,1404.2185} 
for a small subset.
Typical ingredients include the use of the Sachs optical equations 
\cite{Sachs:1961zz} from which a formula for the angular diameter distance 
$d_A$ can be derived,
and approximations of the Dyer-Roeder type \cite{Dyer:1973zz}.
A somewhat different approach is pursued in 
Refs.~\cite{1202.1247,1207.1286,BenDayan:2013gc,Fanizza:2015swa} and related 
papers, where a tailor-made coordinate system \cite{1104.1167} is used.

The present work will take the Sachs optical equations as a starting point, 
but will use them to analyse the evolution of the ``structure distance''
(cf.~Weinberg \cite{Weinberg:2008zzc}) $d_S=(1+z)d_A$.
The result, a second order ordinary differential equation, looks more 
complicated at first sight than the corresponding formula for $d_A$, 
but it turns out that the two nontrivial coefficients have very simple
interpretations: one of them is a local (and directed) expansion rate
that agrees with the standard Hubble rate in the homogeneous case,
and the other one is a quantity that
%direct indicator of inhomogeneity in the sense that it 
vanishes in a spatially flat homogeneous geometry.
These expressions (more precisely: their suitably defined expectation values)
are then computed non-perturbatively in the framework of a recently introduced 
statistical model \cite{1407.6602} whose only assumptions are an irrotational 
dust approximation for the matter content and initial conditions consistent with
linear perturbation theory with only Gaussian fluctuations.
With the help of some approximations (but not of the Dyer-Roeder type) and the
use of a computer program we find that in such a universe with $\L=0$ there
is a time $t_o$ with the following properties. 
An observer at $t_o$ will see redshift-distance pairs
which, if interpreted with formulas that ignore the inhomogeneities, would
indicate $H(t_o)t_o \approx 1$, a deceleration parameter $q(t_o) \approx -0.5$,
and density perturbations at a redshift of $z\approx 1090$ from $t_o$
that agree with those assumed for dark matter at last scattering.
In other words, such an observer sees what present day cosmologists see, 
despite living in a universe in which the cosmological constant vanishes.

In the next section we derive a differential equation for the structure 
distance and discuss the meaning of its coefficients;
furthermore we elucidate the relationship between local expansion data along a
lightlike geodesic and the inferences that a cosmologist who ignores the 
inhomogeneities would make.
In Sec.~\ref{homirrot} the coefficients are computed explicitly for the
cases of homogeneous and irrotational dust universes.
Sec.~\ref{mwa} contains a brief summary of the methods of 
Ref.~\cite{1407.6602} for a non-perturbative statistical treatment of an
irrotational dust universe with initial conditions from linear
perturbation theory.
In Sec.~\ref{secppav} the ``photon path average'' is introduced: this is
the concept that we use to estimate the overall effect of the changing 
environments that a photon experiences on the way from its source to an 
observer.
Sec.~\ref{pertres} contains calculations up to second order in
perturbation theory (we will see that they do not suffice to produce the
relevant effects).
In Sec.~\ref{npres} we present the results of %that were produced by perfoming 
a numerical computation that transcends perturbation theory: 
we find quantities that are in rough agreement with today's observations
even though we assume $\L=0$.
In the final section we briefly reiterate our findings and summarize
the approximations that were made in deriving them.
We also explain why some of the approximations are not as good as they
originally appeared, thus leaving the question of the non-perturbative impact
of inhomogeneities still open; this is the main modification compared to
previous versions of the paper.
%, and discuss some discrepancies with existing literature as well as
% possible directions of future work.

\newpage
\section{Sachs equations and distance formulas}\label{sedf}

Let us start with a brief summary of the homogeneous case in order to provide
some reference points for our subsequent generalization.
A homogeneous universe is usually described with the help of a time-dependent
scale factor $a(t)$ in terms of which the Hubble expansion rate is defined as
\beq H(t) = {\dot a(t) \0 a(t)}, \eeq
and the deceleration parameter as
\beq q = - {\ddot a \, a \0 \dot a^2} = {d\0 dt}\({1\0 H}\) - 1. \eeq
The redshift $z$ of a photon emitted at time $t$ and observed at time $t_o$,
with both the source and the observer at rest with respect to a comoving frame,
is given by 
\beq 1+z = {a(t_o)\0 a(t)}, \eeq
which implies 
\beq H(t) = - {d\0 dt}\ln(1+z). \eeql{Hhom}
%here and elsewhere $t$ indicates the emission time $t_e$.
In the case of vanishing spatial curvature several distance formulas can
be summarized as
\beq d = (1+z)^{\l} \int_0^z{1\0 H(z')} dz', \eeql{distform}
where we have to take $\l=-1$ %if $d$ stands 
for the angular diameter 
distance $d_A$, and $\l=1$ for %if $d$ is 
the luminosity distance $d_L$.
The resulting identity $d_L/d_A=(1+z)^2$ actually holds in any 
pseudo-Riemannian geometry; this is known as Etherington's theorem
\cite{Gen.Rel.Grav.39:1055}.
The simplest version of Eq.~(\ref{distform}) occurs if we take $d$ to be 
the geometric mean of $d_A$ and $d_L$,
\beq d_S = (1+z)d_A  = (1+z)^{-1}d_L, \eeql{dS}
for which there exists a variety of names in the literature;
we will follow Weinberg  \cite{Weinberg:2008zzc} who calls $d_S$ the 
``structure distance''.
Then $\l= 0$, and Eq.~(\ref{distform}) implies
\beq H = {dz\0 dd_S}\eeql{Hzds}
and, with Eq.~(\ref{Hhom}),
\beq dd_S = -(1+z) dt.\eeql{ddSdt}

In the following we consider an arbitrary spacetime geometry.
We want to analyse a light-like geodesic corresponding to the path of a photon
emitted at $x^\m_e$ and observed at $x_o^\m$.
With an affine parameter $s$ and a corresponding tangent vector 
$k^\m = dx^\m/ds$ %=(1+z)\dot x^\m$ 
the redshift $z$ is determined in general by the formula
\beq 1+z = {(u\cdot k)_e\0 (u\cdot k)_o},\eeq
where $u_e$ and $u_o$ are the normalized tangent vectors to the worldlines of
the source and the observer, respectively.
If we assume that we have a distinguished timelike coordinate $t$ such that
both the source and the observer have worldlines with normalized tangent 
vectors $\6 /\6 t$, and that $s$ is normalized so that $ds=dt$ at the 
observer, we get
\beq 
1+z = {dt\0ds}, \quad \hbox{i.e.} \quad {d\0ds} = (1+z){d\0dt}
\eeql{redshift}
(to be evaluated at the source, i.e.~at $t = t_e$; the same holds for 
the following equations).
%Using dots to or write ${\partial\0 \partial t}$ when we differentiate 
%by the spacetime coordinate $t = x^0$ but 
We write ${d\0 dt}$ or use dots when we treat $t$ as parametrizing the 
geodesic, and we denote the partial derivative by the spacetime coordinate
$t = x^0$ as $\6_0$ or ${\partial\0 \partial t}$.

The Sachs optical equations \cite{Sachs:1961zz}
(see \cite{Straumann} for a textbook derivation) are
\bea
- {d \topt \0 ds} + \topt^2 + |\sopt|^2 &=& - \2 R_{\a\b}k^\a k^\b ,
\label{topt}\\
 -{d\sopt\0 ds} + 2 \topt\sopt &=& - \2 R_{\a\b\m\n}\e^\a k^\b\e^\m k^\n,   
\label{sachs2}
\eea
where $\topt$ and $\sopt$ are the expansion rate and the shear of the null 
bundle, respectively.
In general the terms expansion rate and shear refer to the change in the
size and the shape of a bundle of geodesics.
Since we will later apply the same notions to worldlines of dust particles,
we indicate with the subscript that we are referring to the optical quantities.
Furthermore
$\e = \e_{(1)}+ \sqrt{-1}\,\e_{(2)}$ where $\e_{(1)}$, $\e_{(2)}$
are spacelike unit vectors orthogonal both to $k$ and to the observer's 
worldline;
because of these properties the right-hand side of the second equation 
remains the same if the Riemann tensor $R_{\a\b\m\n}$ is replaced by 
the Weyl tensor $C_{\a\b\m\n}$, and corresponding  effects 
%resulting from Eq.~(\ref{sachs2})
are often referred to as ``Weyl focusing''.
The angular diameter distance $d_A$ is determined by
\beq -{d\0 ds} \ln d_A = \topt, \eeq
%\qquad \iff \qquad - {d\0 dt} \ln d_A = (1+z)^{-1}\topt
which can be used to reformulate the Sachs equations as
\bea 
{d^2d_A \0 ds^2} &=& - (|\sopt|^2 + \2 R_{\a\b}k^\a k^\b) d_A, \label{dA}\\
{d\0 ds}(\sopt d_A^2) &=& \2 R_{\a\b\m\n}\e^\a k^\b\e^\m k^\n d_A^2.\label{opts}
\eea
%For the present study we find it most convenient to work with yet another
%closely related distance measure
%``square root of $d_Ld_A$.''
We now want to transform Eq.~(\ref{dA}) into an equation for the structure 
distance $d_S = (1+z)d_A$ as a function of time.
By using Eq.~(\ref{redshift}) we find 
%and (\ref{dS}) 
\beq \ddot d_S -[\ln(1+z)]\rdot \,\dot d_S + id_S = 0\eeql{dSeq}
with 
\beq i = (1+z)^{-2}(|\sopt|^2 + \2 R_{\a\b}k^\a k^\b)-{d^2\0 dt^2}\ln(1+z). 
\eeql{ieq}
%In the case of $i=0$, which holds for spatially flat homogeneous universes,
%The point about this definition is the fact that, a
As we will demonstrate in Sec.~\ref{homirrot}, the quantity $i$ actually 
vanishes for spatially flat homogeneous universes.
In that case Eq.~(\ref{dSeq}) is solved by
\beq \dS0 = \int_{t_e}^{t_o} (1+z)dt = \int_0^z {1\0 -[\ln(1+z)]\rdot} dz.
\eeql{dS0}
%In the general case 
Even for $i\ne 0$ the introduction of $\dS0$ is useful because we can 
simplify Eq.~(\ref{dSeq}) by treating 
$d_S$ as a function of $\dS0$, which results in
\beq {d^2 d_S \0 d\dS0^2} = {-i\0 (1+z)^2} d_S \eeql{dSdS0}
with boundary conditions at $\dS0 = 0$ given by
\beq d_S = 0, \quad {d d_S \0 d\dS0} =1. \eeql {dSdS0bound}

There is no perfectly natural way of generalizing the concept of a Hubble 
rate to an inhomogeneous universe.
Two operational definitions of a ``Hubble rate'' associated with a specific 
point on a geodesic can be made as generalizations of Eq.~(\ref{Hzds}):
\beq \Hd = {d z\0 d d_S}, \qquad \Ht = {d z\0 d \dS0}.\eeq
Both formulas reduce to the standard Hubble rate 
for the case of a homogeneous spatially flat universe.
\del
 as %for $\Hd$ this is 
a consequence of the well-known formula 
$ d_A = (1+z)^{-1} \int^z H^{-1}(z') dz'$.
\enddel
While $\Hd$ is essentially %closely related to
the quantity that is inferred from observations under the assumption
of flat homogeneity,
$ \Ht$ is the expansion at the source in the direction of the
photon emission: by virtue of Eq.~(\ref{dS0}) we have
\beq \Ht = -{d\0 dt}\ln(1+z), \eeql{Htz}
in perfect analogy with Eq.~(\ref{Hhom});
also note that $\Ht$ is just the second coefficient in Eq.~(\ref{dSeq}).
With the help of Eqs.~(\ref{dSdS0}) and (\ref{dSdS0bound}) we find
\beq {\Ht \0 \Hd} ={d d_S\0 d \dS0} 
    =  1 + \int_0^{\dS0} {-i\0 (1+z)^2} d_S \,d\dS0\!'
    =  1 - \int_{t}^{t_o} {i\0 (1+z)} d_S \,dt'. \eeql{HtHd}
This means that the two definitions of $H$ coincide at the observer,
$\Ht(t_o) = \Hd(t_o) = H_o$, and that
for positive $i$ observations tend to overestimate
and for negative $i$ to underestimate expansion rates in previous epochs;
in particular, for sufficiently large negative $i$ we can perceive
acceleration even if it does not take place.

As we have seen, someone who ignores the nonvanishing of $i$ (in other words, 
any cosmologist believing in the standard concordance model) would
interpret $\Hd$ as ``the Hubble rate''. Furthermore, from Eq.~(\ref{ddSdt})
%since Eq.~(\ref{dS0}) implies $dt = -d \dS0/(1+z)$, and $i=0$ would mean 
%$d_S = \dS0$, 
such a person would (wrongly!) infer a time parameter 
$t_\mathrm{inf}$ with
\beq dt_\mathrm{inf} = -{d d_S \0 1+z} = -{\dot d_S \0 1+z}dt.\eeq
In fact, $\Hd$ and $t_\mathrm{inf}$ satisfy an analogue of Eqs.~(\ref{Hhom})
and (\ref{Htz}):
\beq \Hd = {dz\0 d d_S} = -(1+z) {dz\0 d t_\mathrm{inf}} 
       = -{d\0 dt_\mathrm{inf}}\ln(1+z). \eeq
Let us also introduce the deceleration parameters
\beq q_\mathrm{inf} = {d\0 dt_\mathrm{inf}}\({1\0 \Hd}\) -1,\quad
q_\sharp = {d\0 dt}\({1\0 \Ht}\) -1.  \eeq
By using the chain rule, the definitions of the various quantities and 
Eq.~(\ref{dSeq}) one can show that they are related via
\beq q_\mathrm{inf} = q_\sharp + i\,{d_S(1+z)\0 \dot d_S~\dot z}. \eeql{qq}
This demonstrates again that negative $i$ can lead to the perception of 
acceleration even if it does not take place.

We can summarize the results of this section in the following way.
From the values of the pairs $(d_S, z)$ along a given lightlike geodesic, 
without taking into account the quantity $i$ that encodes
the effects of curvature and inhomogeneity, one would infer an expansion
history along that geodesic in terms of quantities $t_\mathrm{inf}$,
$\Hd$ and $q_\mathrm{inf}$.
The actual expansion history \emph{along that specific geodesic} is encoded by
$t$, $\Ht$ and $q_\sharp$.
The two sets of quantities are related by Eqs.~(\ref{HtHd}), (\ref{qq}) and
\beq\Hd \,dd_S = \Ht \,dd_{S\sharp} = dz,\eeq
\beq\Hd \,dt_\mathrm{inf} = \Ht \,dt = -d\ln(1+z).\eeq
In reality we have at most a single data point $(d_S, z)$ for any observed
direction, and we require a statistical analysis.
As we will see, even $\Ht$ and $q_\sharp$ (suitably averaged over photon paths)
can become quite different from the corresponding results from %of some 
volume averaging. % scheme.

\section{Homogeneous and irrotational dust universes}\label{homirrot}

While all of our results up to now are exact in an arbitrary 
pseudo-Riemannian geometry with a distinguished timelike coordinate,
we assume in the following 
that the metric can be written, in the synchronous gauge, as
\beq ds^2 = g_{\a\b} dx^\a dx^\b= -dt^2 + g_{ij}(t,x) dx^i dx^j; 
\eeql{metric}
this is true for any homogeneous spacetime as well as for irrotational dust,
where the dust particles have constant space coordinates $x^i$.
We want to express our quantities in terms of the
spatial 3-geometry with the time-dependent metric $g_{ij}$.
To distinguish it from the spacetime geometry we adopt the convention
that an expression with greek indices or at least one index of zero or a left
superscript of ${}^{(4)}$ pertains to the 4-metric $g_{\a\b}$, 
whereas any other quantity, in particular the Ricci scalar $R = R_i^i$, 
refers to %the spatial 3-metric 
$g_{ij}$.
The connection coefficients for the metric (\ref{metric}) vanish if two or 
three indices are 0, 
and the non-vanishing coefficients are %Christoffels are determined by
\beq \G_{0ij} = -\2 \6_0g_{ij},\quad \G_{i0j} = \G_{ij0} = \2 \6_0g_{ij},\quad 
{}^{(4)}\!\G_{ijk} = \G_{ijk}, \eeq
with the notation $\6_0$ for $\6/\6 x^0 = \6/\6 t$ and more generally
$\6_\m$ for $\6/\6 x^\m$, %in particular 
so that 
\beq {d\0 dt}  = \6_0  + \dot x^i \6_i. \eeql{ddt}
The expansion tensor $\th^i_j$ and  the scalar expansion rate $\th$
are defined by
\beq
\th^i_j=\2 g^{ik}\6_0 g_{kj}, \qquad \th = \th^i_i = {\6_0\sqrt{g}\0\sqrt{g}},
\eeql{exptens}
and the shear is the traceless part of the expansion tensor,
\beq\s^i_j = \th^i_j - {1\0 3}\th \d^i_j,\qquad \s^2 = \2 \s^i_j\s^j_i.\eeq
The Riemann tensor $R_{\a\b\g\d}$ can be expressed in terms of the expansion 
tensor and the Riemann tensor $ R_{ijkl}$ of the spatial metric $g_{ij}$:
\bea 
R_{0i0j} &=& -g_{ik}\6_0\th^k_j - \th_{ik} \th^k_j ,\\
%= -g_{ij}\({\dot \th \0 3} + {\th^2\0 9}\) -g_{ik}\dot\s^k_j 
%- {2\0 3} \th \s_{ij} - 2 \s_i^k \s_k^l g_{lj},\\
R_{0ijk} &=& \th_{ij|k} - \th_{ik|j}, \\
{}^{(4)}\! R_{ijkl} &=&  R_{ijkl} - \th_{il}\th_{jk} + \th_{ik}\th_{jl},
\label{Riem4}
\eea
with $\th_{ij} = g_{ik}\th^k_j$ and with the vertical strokes denoting 
covariant spatial derivatives.

We now want to specialize our analysis of photon paths to a metric of the type
(\ref{metric}), with the assumption that both the source and the observer
are comoving: $x^i_e = \mathrm{const}$, $x^i_o = \mathrm{const}$.
Since $\G^0_{ij}= \2 \6_0 g_{ij}$, the 0-component of the 
geodesic equation is
%for $t = x^0$ becomes
\beq {d^2 t\0 ds^2} + \2 (\6_0 g_{ij}) {dx^i\0 ds}{dx^j\0 ds} = 0\eeq
or, upon division by $(1+z)^2$ and application of Eq.~(\ref{redshift}), 
\beq {1\0(1+z)^2}{d(1+z)\0 ds} = -\2 (\6_0 g_{ij}) \dot x^i \dot x^j.\eeq
As $\dot x^\m$ is light-like and $x^0 = t$, the spatial part $\dot x^i$ 
must be a unit vector with respect to $g_{ij}$, 
\beq g_{ij}\dot x^i \dot x^j = 1, \eeql{xdot}
whereby the previous equation becomes
\beq {d\0 dt} \ln(1+z) = - {\th\0 3} - \s_{ij}  \dot x^i \dot x^j.
\label{lndot}\eeq
Similarly we can transform the spatial component
\beq 
{d^2 x^i\0 ds^2} + 2 \th^i_j {dt\0 ds}{dx^j\0 ds} + 
\G^i_{jk}{dx^j\0 ds}{dx^k\0 ds} = 0
\eeq
of the geodesic equation into
\beq 
\ddot x^i + {\th \0 3} \dot x^i - \s_{kl} \dot x^k \dot x^l \dot x^i
  + 2 \s^i_j \dot x^j + \G^i_{jk}\dot x^j\dot x^k = 0.
\eeq
Upon using this, together with (\ref{ddt}), in the derivative of 
Eq.~(\ref{lndot}), we find
\beq -{d^2\0 dt^2}\ln(1+z) = (\6_0 +\dot x^i \6_i){\th \0 3}
  + (\6_0 \s_{ij} + \dot x^k \6_k \s_{ij}) \dot x^i \dot x^j
  - 2 \s_{ij}({\th \0 3} \dot x^i - \s_{kl} \dot x^k \dot x^l \dot x^i
  + 2 \s^i_k \dot x^k + \G^i_{kl}\dot x^k\dot x^l)\dot x^j.
\eeql{mldd}
%
%If we assume validity of 
Note that up to now we have never used the Einstein equations
\beq R_{\a\b}-\(\2\,{}^{(4)}\!R - \L\)g_{\a\b} = 8 \pi G_N T_{\a\b}. \eeql{Einst}
Let us assume that the spatial part of the energy-momentum tensor is 
proportional to the metric, $T_{ij} = g_{ij}T_k^k/3$, and that $T_{0i} = 0$.
This holds not only in the homogeneous case but also in the general
irrotational dust case, where $T_{ij} = 0$.
Then Eq.~(\ref{Einst}) implies that the spacetime Ricci tensor $R_{\a\b}$
must be of the same type, 
$ {}^{(4)}\!R_{ij} = g_{ij}\,{}^{(4)}\!R_k^k/3$ and $R_{0i} = 0$, so that
\beq
R_{\a\b}k^\a k^\b =  R_{00}(k^0)^2 + {1\0 3} g_{ij}k^ik^j\,{}^{(4)}\!R_k^k
= (1+z)^2 (R_{00}+{1\0 3} \,{}^{(4)}\!R_k^k);\eeq
in the last step we have used $k^0 = dx^0/ds = 1+z$ and 
$g_{ij}k^ik^j= k_\m k^\m + (k^0)^2 = (1+z)^2$.
With the help of Eqs.~(\ref{exptens}) -- (\ref{Riem4}) this results in
\beq \2 (1+z)^{-2} R_{\a\b}k^\a k^\b = - {1\0 3} \6_0\th + {R\0 6} - \s^2.
\eeql{hRkk}
The traceless spatial part of the Einstein equations amounts to
\beq \6_0\s^i_j + \th \s^i_j + r^i_j = 0, \eeq
which implies $\6_0 \s_{ij} = - \th \s_{ij} / 3 + 2 \s_i^k \s_{kj} - r_{ij}$,
where 
\beq r_{ij}= R_{ij} - {R\0 3}g_{ij} \eeq 
represents the traceless part of the spatial Ricci tensor.
Using this after inserting Eqs.~(\ref{mldd}) and (\ref{hRkk}) 
into (\ref{ieq}) we get
\bea
i &=& (1+z)^{-2}|\sopt|^2 + R/6 - \s^2 
    + (- \s_{ij}\th - 2 \s_i^k \s_{kj} - r_{ij}
            + 2 \s_{ij}\s_{kl} \dot x^k \dot x^l )\dot x^i\dot x^j   \nn\\
   && + \dot x^i \6_i\th / 3
  + \dot x^k (\6_k \s_{ij}) \dot x^i \dot x^j
  - 2 \s_{ij}\G^i_{kl}\dot x^k\dot x^l\dot x^j.
   %~(\hbox{terms that are odd in } \dot x).
\label{iirrd}\eea
This result is still exact within the irrotational dust framework and also 
for any homogeneous cosmological model. 
In the latter case it reduces to $i=R/6 = K/a^2$ with $K\in\{-1,0,1\}$
%$R$ proportional to $(a_o/a)^2 = (1+z)^2$
so that $i/(1+z)^2 = K/a^2_o$ is constant; 
thereby Eqs.~(\ref{dS0}), (\ref{dSdS0}) lead 
to the well known distance formulas that involve sin or sinh functions
%solutions of Eq.~(\ref{dSdS0}) for $R \ne 0$ and to $d_S = \dS0$ 
for $K \ne 0$.

Let us also note that the equation (\ref{opts}) for the optical shear is 
determined by
\beq R_{\a\b\m\n}\e^{\a}k^\b\e^{\m}k^\n
  = (1+z)^2({2\0 3} \th\s_{ij}-\s_{ik}\s^k_j+2r_{ij} 
    + \dot x^l\s_{lm}\dot x^m\s_{ij}
    -\dot x^l\s_{li}\dot x^m\s_{mj} - 4\dot x^k\s_{i[j|k]})\e^i\e^j
\eeql{optsrhs}
%in the present %irrotational dust scenario.
for any metric of the type (\ref{metric}), as one can ascertain by using 
similar methods.
This expression vanishes for any homogeneous model.

\newpage
\section{Mass-weighted average}\label{mwa}

If we knew the spatial metric $g_{ij}$ in the vicinity of a given
lightlike geodesic in an irrotational dust universe, we could now 
compute the redshift and the structure distance along that geodesic simply by
solving Eqs.~(\ref{lndot}) and (\ref{dSeq}) %or (\ref{ddsdt}) 
with input
from Eq.~(\ref{iirrd}) (assuming we are also solving for $\sopt$ along the way).
In practice we do not know the precise form of the metric and need to rely 
on a statistical model; 
in addition we have to make simplifications to keep the computations manageable.
As we aim for results beyond perturbation theory, we choose the approach of 
Ref.~\cite{1407.6602} for our underlying statistical model.
The present section is devoted to a brief summary of the relevant ideas
and results.
The central concept in this approach is the mass-weighted average \cite{1310.1028}
\beq 
\< X\>_\mathrm{mw}(t) = {1\0 m_\cd} \int_\cd X(x,t)\rho(x,t)\sqrt{g(x,t)} ~d^3 x
\eeql{xavg}
of a scalar quantity $X$,
where $\cd$ is a large domain (e.g.~all of the visible universe),
$\rho(x,t)$ is the local mass density and 
\beq m_\cd = \int_\cd \rho(x,t)\sqrt{g(x,t)} ~d^3 x  \eeql{mc}
is the mass content of $\cd$.
For the case of an irrotational dust universe, energy conservation implies
\beq {\6\0 \6 t} \(\rho(x,t)\sqrt{g(x,t)}\) = 0 \eeq
and therefore $\< \6_0 X\>_\mathrm{mw} = \6_0\< X\>_\mathrm{mw}$. 
This makes it possible to evade the technical difficulties that arise with
the more common volume average, where averaging and taking time 
derivatives do not commute.
Nevertheless volume averages are easily computed within this approach as
\beq \< X\>_\mathrm{vol}
= {\< X \rho^{-1}\>_\mathrm{mw}\0\<\rho^{-1}\>_\mathrm{mw}}
= {\< X a^3\>_\mathrm{mw}\0 \<a^3\>_\mathrm{mw}};
\eeq
here $a$ is the local scale factor defined as
\beq a(t,x) = \({\hat\rho \0 \rho(t,x)}\)^{1\0 3},\eeq
where $\hat\rho$ is an arbitrary fixed mass. 
Then the dust expansion rate can be expressed as
\beq  \th(t,x) = - {\6_0 \r(t,x) \0  \r(t,x)} = 
3 {\6_0 a(t,x) \0  a(t,x)}, \eeq
and a set of rescaled quantities
\beq \hat\rho = a^3 \rho,  ~~~ \hat\s^i_j = a^3 \s^i_j,  
~~~ \hat R = a^2~ R,    ~~~ \hat r^i_j = a^2 r^i_j \eeq
obeys the evolution equations
\beq
\6_0{\hat\rho} = 0,\quad
{\6_0{\hat\s}^i_j} = - a \hat r^i_j,\quad
\6_0{\hat R} = -2 a^{-3} \hat\s^i_j \hat r^j_i,
\eeql{eveq}
\beq \6_0{\hat r^i_j} =   
   a^{-3}\(-{5\0 4} \hat\s^i_k\hat r^k_j +  {3\0 4} \hat\s^k_j\hat r^i_k 
   + {1\0 6} \d^i_j \hat\s^k_l\hat r^l_k\) + a^2{Y^{ki}}_{j|k},\label{rhatev}
\eeq
where
\beq 
{Y^k}_{ij} = {3\0 4} (\s^k_{i|j}+\s^k_{j|i})-\2g_{ij}{\s^k_{m|}}^m-{\s_{ij|}}^k.
\eeq
The initial values for these evolution equations can be found by 
comparison with linear perturbation theory:
upon neglecting vector, tensor and decaying scalar modes the space metric 
$ g_{ij}^\mathrm{(LPT)}(t,x)$ at early times can be expressed in terms of
a single time-independent scalar Gaussian random function $C(x)$ as
\beq g_{ij}^\mathrm{(LPT)}(t,x) = a_\mathrm{EdS}^2(t)
  \(\d_{ij} + {10\0 9}{a_\mathrm{EdS}^2\0 t^{4\0 3}}C(x)\d_{ij} 
   + t^{2\0 3} \partial_i\partial_jC(x)\);   \eeql{gLPT}
here $a_\mathrm{EdS} = \mathrm{const} \times t^{2/3}$ is the standard EdS
(Einstein-de Sitter, i.e.~flat matter-only FLRW) scale factor.
% for the spatially flat case.
By comparing with section 5.3 of Ref.~\cite{Weinberg:2008zzc} one finds that 
this metric is equivalent to a Newtonian gauge metric with $\Phi = \Psi = -C/3$.
It turns out that the initial conditions for our evolution equations are
%determined by the second derivatives of $C$:
\bea 
\lim_{t\to 0}\,{a\0 t^{2\0 3}} &=& (6\pi G_N\hat \rho)^{1/3},\\
\hat \s_{\mathrm{in}}(x) &=& 0,\\
\hat R_{\mathrm{in}}(x) &=& -  
{20\0 9}(6\pi G_N\hat \rho)^{2\0 3}S(x),\\
({\hat r}_{\mathrm{in}})^i_j(x) &=& 
-  {5\0 9}(6\pi G_N\hat \rho)^{2\0 3}\d^{ik}s_{kj}(x),
\eea
where $S$ and $s_{kj}$ are the trace and traceless parts of the matrix
\beq \partial_i\partial_jC(x)=S_{ij}(x)=s_{ij}(x)+{1\0 3}\d_{ij}S(x)\eeq
of second derivatives of the function $C(x)$.
In this setup it can be shown that
\beq
\hat R(t) = \hat R_{\mathrm{in}} + 2 a^{-4}(t)\,\hat\s^2(t) 
+ {8\0 3}\int_{t_{\mathrm{in}}}^t \th(\tilde t)a^{-4}(\tilde t)\,
\hat\s^2(\tilde t)\,d\tilde t,
\eeql{Rhat}
and that the evolution equation of the local scale factor $a(x,t)$ is
\beq 
(\6_0 a)^2  = {8\0 3}\pi G_N\hat\rho \,a^{-1}
- {1\0 6}\hat R_\mathrm{in} + {1\0 3}\L\,a^2 - {4\0 9}
\int_{t_{\mathrm{in}}}^t\th(\tilde t)a^{-4}(\tilde t)\,\hat\s^2(\tilde t)\,d\tilde t.
\label{lsfev}\eeq
As long as one neglects the last term ($a^2{Y^{ki}}_{j|k}$) in 
Eq.~(\ref{rhatev}), the evolution in a given region will depend only on the 
initial conditions within that region; furthermore, if one chooses a coordinate
system in which the symmetric matrix $S_{ij}(x)$ is diagonal then $r_{ij}$ and 
$\s_{ij}$ will be diagonal in that system at any time $t$.
In this way it suffices to work with the probability distribution for the three 
eigenvalues of $S_{ij}$.
% which is given 
%This distribution was computed explicitly i
As shown in Ref.~\cite{1407.6602}, %it was shown that 
the assumption that $C(x)$ is a 
Gaussian random field suffices to compute this distribution explicitly in terms
%up to 
%The resulting statistical model depends only on 
of a single dimensionful parameter which is related to the value of an 
integral that requires an ultraviolet cutoff.
%sets the scale.
Then one can switch to dimensionless units by taking a specific value
for this parameter.
With the computationally convenient choice that was
adopted in Ref.~\cite{1407.6602} and that will also be used here, one finds
\beq \<S^2\>_\mathrm{mw} = 5,\quad 
\< s_{ij}s_{kl}\d^{ik}\d^{jl}\>_\mathrm{mw} = 10/3. \eeql{sevs}
If one also chooses $\hat\rho$ such that $6\pi G_N \hat\rho = 1$
in the corresponding units then the perturbative series for $a$ starts as
\beq a(x,t) = t^{2\0 3} + {S(x)\0 6} t^{4\0 3} 
- {S^2(x) + 2  s_{ij}(x)s_{kl}(x)\d^{ik}\d^{jl} \0 84} t^2 + \ldots,
\eeql{apert}
where we have neglected cubic and higher orders in perturbation theory.

In the following we will develop the theory further in terms of the
dimensionless quantities that we have introduced here.
This leads to unique results (up to ambiguities in approximations), with
the only free parameter coming from the reintroduction of a dimensionful
scale once we start comparing our results with physical quantities.

\section{Photon path average}\label{secppav}

Finally we want to connect the distance formula (\ref{dSeq}), which relies
on the values of the quantities $\Ht = -[\ln(1+z)]\rdot$ and $i$ along a
photon path, with the model of Ref.~\cite{1407.6602} as summarized above.
We propose to do the following. 
We replace the right-hand sides of Eqs.~(\ref{lndot}) and (\ref{iirrd})
by suitable expectation values which we will denote by 
$\<~\ldots ~\>_\mathrm{pp}$, where the subscript stands for ``photon path''.
The idea is that $\<X\>_\mathrm{pp}(t)$ should be the average of $X$ over
all spatial positions $\mathbf{x}$ occupied by a photon of a given type
(e.g.~supernova or CMB) at the time $t$, as well as all directions 
$\mathbf{v}$ of propagation of such a photon.
A complete realization of this concept would automatically guarantee 
consistency with correct ensemble and angular averaging.
While the approximation we will make at the beginning of the next paragraph
leads to a mild angular deviation, %all of the quantities we compute 
statistical isotropy and homogeneity will be manifest in all our computations.
Every photon path corresponds to a random walk in the probability
space determined by the six entries of $S_{ij}$ (or, alternatively, three 
eigenvalues and three direction components).
Then $X=\<X\>_\mathrm{pp} + \D X$ with $\<\D X\>_\mathrm{pp} = 0$, and by the
linearity of Eq.~(\ref{dSeq}) the contribution of $\D X$ gets small if
a photon probes different regions of the probability space within 
a short time.

Every photon path corresponds to a curve $\cc$ in $\mathbf{x}$--space
(the $\IR^3$ parametrized by the spatial coordinates $x^1$, $x^2$, $x^3$) 
that ends at $\mathbf{x}_o$.
In the flat homogeneus case these curves are just straight lines.
If the shapes of these curves were not altered by the presence of 
inhomogeneities, then our model would tell us how the basic parameters 
are distributed with respect 
to the euclidean metric $dl^2 = \d_{ij} dx^i dx^j$ along such a curve.
%Our next 
We will make the simple approximation of assuming the same distribution
even in the general case. 
As a next step we want to move on to a description that is based on
physical time rather than euclidean length.
We denote by %$l$ the length along $\cc$ as measured with $\d_{ij}$, and by 
\beq v^i={dx^i\0 dl} = \dot x^i ~{dt\0 dl}\eeql{vi}
the tangent vector to $\cc$ normalized to euclidean unit length, 
i.e.~$\d_{ij}v^iv^j = 1$.
Upon taking the $g$-norm $\sqrt{g_{ij}v^iv^j}$ of $\mathbf{v}$ and using
Eq.~(\ref{xdot}) we find 
\beq dt = \sqrt{g_{ij}v^iv^j} \,dl ,\eeql{dt}
which reflects the fact that the photon flight time 
%requires to traverse a path
is proportional to the traversed distance as measured with the physical
metric $g$.
For any path segment of length $dl$ we average over the three basic 
parameters of the model (indicated by $\<~\ldots~\>_\mathrm{mw}$) and over
all directions $\mathbf{v}$, and weight by the time 
$dt = \sqrt{g_{ij}v^iv^j} \,dl$ spent in such a segment. 
This results in
\beq \< X \>_\mathrm{pp} = 
   {\< \int_{S^2}X \sqrt{g_{ij}v^iv^j} d^2v\>_\mathrm{mw} \0
       \< \int_{S^2}\sqrt{g_{ij}v^iv^j} d^2v\>_\mathrm{mw}},
\eeql{ppav}
where the integrations are taken over the unit sphere 
$S^2 = \{\mathbf{v}: \d_{ij}v^iv^j = 1\}$ in tangent space; 
%run over all vectors $\mathbf{v}$ with $\d_{ij}v^iv^j = 1$; 
if $X$ depends on $\dot x^i$ explicitly, we make
use of
\beq \dot x^i = {v^i\0 \sqrt{g_{ij}v^iv^j}} \eeq
which follows from Eqs.~(\ref{vi}), (\ref{dt}).

Our aim is to compute $\<X \>_\mathrm{pp}$ for the nontrivial coefficients
in Eq.~(\ref{dSeq}), i.e.~for the cases 
$X= -[\ln(1+z)]\rdot$ and $ X=i$. %, which involves 
To this end we require %would have to compute 
integrals over $S^2$ of expressions that are polynomials 
in the $v^i$ except for the occurrence of factors of $\sqrt{g_{ij}v^iv^j}$.
Since exact results would involve elliptic functions we 
%are going to use the following approximation. We 
work in a basis in which the metric is diagonal and write
\beq g_{ij} = \bar g (\d_{ij} + \g_{ij}) \eeq
with
\beq   \bar g = {g_{11} + g_{22} + g_{33} \0 3},\quad
\g_{11} + \g_{22} + \g_{33} = 0. \eeql{sumgamma}
Then %so that 
\beq \(\sqrt{g_{ij}v^iv^j}\)^\l = \(\sqrt{\bar g(1 +\g_{ij}v^iv^j)}\)^\l =
   {\bar g}^{\l / 2}(1 + {\l\0 2}~\g_{ij}v^iv^j + \ldots)
\eeq
on the sphere $\d_{ij}v^iv^j = 1$.
For each term in this expansion
we require only integrals of polynomials in the $v^i$, such as
\bea
&\int_{S^2} (v^i)^{2n}d^2v= 4\pi / (2n+1),\quad
\int_{S^2} (v^1)^2(v^2)^2d^2v= 4\pi / 15, &\label{int1}\\
&\int_{S^2} (v^1)^4(v^2)^2d^2v= 4\pi / 35, \quad
\int_{S^2} (v^1)^2(v^2)^2(v^3)^2d^2v= 4\pi / 105.&
\eea
From now on we simply omit any terms that are of quadratic or higher order 
in the $\g_{ij}$.
While this may look excessively crude, one can check that even in the 
extremal cases of one or two vanishing eigenvalues the error is at most 
around 15\%.
For the integral in the denominator of (\ref{ppav}) this gives,
upon using (\ref{sumgamma}),
\beq 
\int_{S^2} 
\sqrt{g_{ij}v^iv^j} d^2v \approx
4 \pi \sqrt{\bar g}.
\eeql{Iapprox}
\del
because 
\beq \g_{11} + \g_{22} + \g_{33} = 0. \eeql{sumgamma}
\enddel
According to Eq.~(\ref{lndot}), 
$ -[\ln(1+z)]\rdot=  \th /3 + \s_{ij}  \dot x^i \dot x^j$.
Since $\th$ has no direction dependence, 
\beq \int_{S^2}{\th\0 3} \sqrt{g_{ij}v^iv^j} d^2v = 
{\th\0 3}\int_{S^2} 
\sqrt{g_{ij}v^iv^j} d^2v \approx 4 \pi \sqrt{\bar g}{\th\0 3}. \eeq
In evaluating the second term we use the fact that $\s_{ij}$ is diagonal
in the same coordinate system in which $g_{ij}$ is: %; thereby we find
\beq \s_{ij}  \dot x^i \dot x^j \sqrt{g_{ij}v^iv^j}
  =  {\s^k_jg_{ki}  v^i v^j \0 \sqrt{g_{ij}v^iv^j}}
  = \sqrt{\bar g}\(\sum_{i=1}^3\s_i^i (1+\g_{ii}) (v^i)^2\) 
   \(1 - {1\0 2}~\sum_{i=1}^3\g_{ii} (v^i)^2 + \ldots\). 
\eeq
Upon restricting this to terms linear in $\g_{ij}$ and using the 
formulas (\ref{int1}) and (\ref{sumgamma}) we get
\beq
\int_{S^2} \s_{ij}  \dot x^i \dot x^j \sqrt{g_{ij}v^iv^j}d^2v
\approx 
{16\0 15}\pi \sqrt{\bar g}(\s_1^1 \g_{11} + \s_2^2 \g_{22} + \s_3^3 \g_{33}).
\eeql{sxxapp}
%and therefore
Combining our results gives
\beq
   \< -[\ln(1+z)]\rdot \>_\mathrm{pp}
\approx   {\< \sqrt{\bar g}(5\th + 4\s_1^1 \g_{11} + 4\s_2^2 \g_{22} 
                         + 4\s_3^3 \g_{33})\>_\mathrm{mw}\0 
15~\< \sqrt{\bar g}\>_\mathrm{mw}}.
\eeql{Hnonpert}

Next we turn our attention to $\<i\>_\mathrm{pp}$.
Since no direction is singled out, the expressions 
%$\dot x^i \6_i (\th/ 3 + \s_{ij} \dot x^i\dot x^j)$
in the second line of Eq.~(\ref{iirrd}), which are all 
odd under $\dot x^i \to - \dot x^i$, do not contribute after averaging.
The optical shear $\s_\mathrm{opt}$ is determined by Eq.~(\ref{opts}).
The behaviour for small $t_o - t$ is easily found to be
$\s_\mathrm{opt} \approx {1\0 6}  (t_o-t)R_{\a\b\m\n}\e^\a k^\b\e^\m k^\n$,
i.e.~well-behaved and vanishing in the limit $t\to t_o$.
Under a $90^\circ$ rotation $\e_{(1)}\to \e_{(2)}$, $\e_{(2)}\to -\e_{(1)}$ the 
right-hand side of Eq.~(\ref{opts}) changes sign, hence its photon path
average vanishes and the behaviour of $\s_\mathrm{opt}$ resembles a random 
walk around zero.
%, which leads to further suppression.
Near $t=0$ we can use the results of linear perturbation theory as
presented in Sec.~\ref{mwa} to find that the right-hand side of
Eq.~(\ref{optsrhs}) behaves like $t^{-8/3}$, hence that of Eq.~(\ref{opts})
like $t^{-4/3}$.
%Without the suppression from the random walk type behaviour
Naively this would
result in $\sopt \sim t^{-1}$ and a contribution of type $t^{-2/3}$ to
Eq.~(\ref{iirrd}), which is the same power as the leading (second order)
behaviour of the other terms, as we will shortly see; because of the
random walk nature it will however be suppressed.
In the following we will neglect the term $(1+z)^{-2}|\sopt|^2$ in
Eq.~(\ref{iirrd}), but keep in mind that $i$ will receive a moderate
positive correction for intermediate redshift values;
in particular we should remember that this makes our results more
reliable for smaller than for larger redshifts.
According to Eq.~(\ref{Rhat}), %Ref.~\cite{1407.6602},
\beq R = a^{-2}\hat R_\mathrm{in} + 2\s^2
   + {8\0 3}a^{-2}\int_0^t\th(\tilde t) a^2\s^2d\tilde t,  \eeq
where $\hat R_\mathrm{in} = \lim_{t\to 0}a^2\,R$.
The contribution of $(-\s_{ij}\th-r_{ij})\dot x^i \dot x^j$ can be treated like
that of $\s_{ij}\dot x^i \dot x^j$ before, resulting in
\beq
\int_{S^2}(-\s_{ij}\th-r_{ij})\dot x^i \dot x^j 
\sqrt{g_{ij}v^iv^j}d^2v  \approx 
-{16\0 15}\pi \sqrt{\bar g}[(\s_1^1\th + r_1^1) \g_{11} + \ldots].
\eeq
With slightly more work we also find 
\beq
\int_{S^2}-2\s_i^k\s_{kj}\dot x^i \dot x^j 
\sqrt{g_{ij}v^iv^j}d^2v  \approx 
-{8\0 15}\pi \sqrt{\bar g}[(\s_1^1)^2(5 + 4\g_{11}) + \ldots]
\eeq
and
\beq
\int_{S^2}2\s_{ij}\s_{kl}\dot x^i \dot x^j \dot x^k \dot x^l
\sqrt{g_{ij}v^iv^j}d^2v  \approx 
{16\0 105}\pi \sqrt{\bar g}[(\s_1^1)^2(7 + 8\g_{11}) + \ldots].
\eeq
Putting the pieces together we obtain
\beq 
{1\0 4\pi}\int_{S^2} i \sqrt{g_{ij}v^iv^j}d^2v  \approx 
\sqrt{\bar g}\({\hat R_\mathrm{in}\0 6 a^2} 
   + {4\0 9a^2}\int_0^t\th(\tilde t) a^2\s^2d\tilde t
   - {22 \0 15}\s^2
   -{4\0 105}[(7\s_1^1\th + 7 r_1^1 + 6 (\s_1^1)^2) \g_{11} + \ldots]\).
\eeql{ippnum}
Our formulas %for $-[\ln(1+z)]\rdot$ and $i$ 
rely explicitly on the spatial metric $g_{ij}$ in the diagonal basis.
To obtain it from the quantities whose evolution is studied in Sec.~\ref{mwa}
we use
\beq \2 \6_0 \ln g_{11} = \2 g^{11}\6_0 g_{11}  
= \th_1^1 = {\th\0 3} + \s_1^1 = (\ln a)\rdot + \s_1^1  \eeq
which implies 
\beq 
g_{11} (t) 
= \hbox{const} \times a^2 \times \exp\(2\int_0^t \s_1^1(\tilde t) d\tilde t\),
\eeql{g11}
with analogous expressions for $g_{22}$ and $g_{33}$.
Comparison with Eq.~(\ref{gLPT}) shows that the constant must be the same 
in each case, and that setting it to 1 corresponds to a normalization where
$\<a^2\> = a^2_\mathrm{FLRW}$.

%\newpage

\section{Perturbative results}\label{pertres}

Before proceeding to the results of a non-perturbative numerical computation,
let us first assume that we are still so close to the EdS 
%(Einstein--de Sitter, i.e.~FLRW with $a\propto t^{2/3}$)
case that in most regions perturbation theory provides a good approximation.
%In the absence of background curvature both $R$ and $\s$ are of first order
%in perturbation theory. 
We work with the dimensionless quantities described at the end of 
Sec.~\ref{mwa}.
Again our first goal is the photon path average of the right-hand side of
Eq.~(\ref{lndot}).
From Eq.~(\ref{apert}) we find (to the same accuracy as there)
\beq \th(x,t) = 2 t^{-1}\(1 + {S(x)\0 6} t^{2\0 3} 
- {13 S^2(x) + 12  s_{ij}(x)s_{kl}(x)\d^{ik}\d^{jl} \0 252} t^{4\0 3} + \ldots \).
\eeql{thpert}
The approximation (\ref{Iapprox}) is valid at linear order, 
%In order to compute $\<\th I \>_\mathrm{mw}/ \<I \>_\mathrm{mw}$ for 
and with Eq.~(\ref{g11}) and the fact that $\s_i^j$ is traceless we get 
\beq 
I:= \int_{S^2}\sqrt{g_{ij}v^iv^j} d^2v = 4 \pi a + \co(2);
\eeql{Idef}
%we notice that 
$\co(n)$ means an expression of $n^\mathrm{th}$ or higher order in 
perturbation theory.
Since the 
perturbative expansions $\th = \th^{(0)} + \th^{(1)} +  \th^{(2)} + \co(3)$
and  $I = I^{(0)} + I^{(1)} +  I^{(2)} + \co(3)$ have deterministic leading terms
(i.e., $\th^{(0)} = \<\th^{(0)} \>_\mathrm{mw}$ and 
$I^{(0)} = \<I^{(0)} \>_\mathrm{mw}$) and first order terms whose expectation 
values vanish (i.e., $\<\th^{(1)} \>_\mathrm{mw} = 0$ and 
$\<I^{(1)} \>_\mathrm{mw} = 0$), we get
\beq 
\<\th \>_\mathrm{pp} = { \<\th I \>_\mathrm{mw}\0 \<I \>_\mathrm{mw}} 
= \th^{(0)} + \< \th^{(2)} + {\th^{(1)} I^{(1)} \0  I^{(0)}}\>_\mathrm{mw} + \co(3);
\eeq
note that $I^{(2)}$ has dropped out at quadratic order so that 
Eqs.~(\ref{sevs}), (\ref{apert}), (\ref{thpert}) and (\ref{Idef}) 
suffice for computing
\beq \<\th \>_\mathrm{pp} \approx 2 t^{-1} - {5\0 9} t^{1\0 3} \eeql{thpp}
to the same order as $a$ and $\th$ before.
%For $\< \s_{ij}\dot x^i \dot x^j\>_\mathrm{pp}$ we start with the 
%observation that t
The approximation (\ref{sxxapp}) implies
\beq \< \s_{ij}\dot x^i \dot x^j\>_\mathrm{pp} 
\approx {4\0 15}\< \s_1^1 \g_{11} + \s_2^2 \g_{22} + \s_3^3 \g_{33}\>_\mathrm{mw}
\eeq
at leading (second) order.
This can be evaluated via
\beq
\s_1^1 = - a^{-3}\int_0^t a\, \hat r_1^1 \,d\tilde t  
\approx -{3\0 5}t^{-{1\03}} \,\hat r_1^1 \approx {1\0 3} t^{-{1\03}}s_{11} 
\eeql{sig11app}
(here and in the following equation we only consider leading orders),
\beq \g_{11} = {g_{11} \0 \bar g} -1 \approx 
e^{2\int_0^t \s_1^1 d\tilde t} - 1 \approx 
2\int_0^t \s_1^1 d\tilde t \approx t^{2\0 3} s_{11}
\eeql{gam11app}
and Eq.~(\ref{sevs}); the result is 
\beq \< \s_{ij}\dot x^i \dot x^j\>_\mathrm{pp} \approx {8\0 27} t^{1\0 3}.
\eeq
Combining this with Eq.~(\ref{thpp}) we obtain
\del
Since $\s_i^j$ is traceless and of first order in perturbation theory,
Eq.~(\ref{g11}) implies
$\bar g = a^2 + \co(2)$, and with Eq.~(\ref{apert}) and 
$\< S\>_\mathrm{mw} = 0$ we get
\beq \<\sqrt{\bar g}\>_\mathrm{mw} = t^{2/3} + \co(2). \eeq
Similarly we find 
$  \<\th \sqrt{\bar g}\>_\mathrm{mw} = 2 t^{-1/3} + \co(2)$ and
$\s_1^1 \g_{11} + \s_2^2 \g_{22} + \s_3^3 \g_{33} = \co(2)$
so that 
\enddel
\beq 
\< \Ht\>_\mathrm{pp} \approx {2\0 3} t^{-1} + {1\0 9} t^{1\0 3}, 
\eeql{Hpert}
where the approximation again neglects terms of cubic or higher order in
perturbation theory.
%as well as terms with higher derivatives of $C(x)$.

In order to compute $\< i \>_\mathrm{pp}$ up to second order in perturbation 
theory we require the mass-weighted average of Eq.~(\ref{ippnum}).
We begin with
\beq 
\< \sqrt{\bar g}{\hat R_\mathrm{in}\0 a^2}\>_\mathrm{mw}
 \approx \< {\hat R_\mathrm{in}\0 a}\>_\mathrm{mw}
 \approx -{20\0 9}t^{-{2\0 3}}\< S(1-{1\0 6}t^{2\0 3}S)\>_\mathrm{mw}
= {10\0 27} \< S^2\>_\mathrm{mw} = {50\0 27},
\eeq
where the approximations neglect contributions 
of third or higher order in perturbation theory;
the linear term has dropped out upon averaging.
All other expressions in Eq.~(\ref{ippnum})
%contributions to $\<i\sqrt{\bar g}\>_\mathrm{mw}$ 
are explicitly of quadratic or higher order:
with Eq.~(\ref{sig11app}) we find
\beq \s^2 \approx {1\0 18} t^{-{2\03}}(s_{11}^2 + \ldots),\eeq 
\beq {1\0 a^2}\int_0^t\th a^2\s^2d\tilde t %\approx 3 \s^2 
\approx {1\0 6} t^{-{2\03}}(s_{11}^2 + \ldots),\eeq 
and Eq.~(\ref{gam11app}) together with
\beq  r_1^1 = a^{-2} \hat r_1^1 \approx - {5\0 9} t^{-{4\0 3}}s_{11} \eeq
implies
\beq 
(\s_1^1\th +  r_1^1)\g_{11} + \ldots 
\approx ({2\0 3} - {5\0 9}) t^{-{4\0 3}} t^{2\0 3} (s_{11}^2 + \ldots)
= {1\0 9}t^{-{2\0 3}}(s_{11}^2 + \ldots).
\eeq
Combining all contributions and using Eq.~(\ref{sevs}) we arrive at
\beq 
\< i\>_{\mathrm{pp}} \approx 
\({1\0 6}\times{50\0 27} + ({4\0 9}\times {1\0 6}- {22\0 15}\times{1\0 18}
-{4\0 15}\times {1\0 9}){10\0 3}\)t^{-{2\03}}
= {5\0 27}t^{-{2\03}}.
\eeq

\newpage

\section{Non-perturbative results}\label{npres}
In this section we present the results of numerical computations
performed with GNU octave \cite{octave}.
We used the Euler method with logarithmic time steps to solve the evolution
equations (\ref{eveq}) and (\ref{lsfev}).
We assumed, however, constant $\hat r = \hat r_\mathrm{in}$ instead of 
using Eq.~(\ref{rhatev}), for the following reasons:
the last term in that equation describes wavelike perturbations which
probably play no role and cannot be described directly within the
present model, and 
the other terms have extremely little impact on overall results
(at least when volume evolutions are studied, see  Fig.~12 of
Ref.~\cite{1407.6602} and note that the tiny deviations only occur for
$t\gg 1$).
This was done for a large set of initial conditions, 
and the resulting values for $a$, $\s$, $r$ and $R$ were used to evaluate the 
formulas of Sec.~\ref{secppav}, with an appropriate probability measure
for each set of initital conditions.
More algorithmic details can be found in
the appendix of Ref.~\cite{1407.6602}. %; to be specific, w

In regions that collapse, the treatment in terms of irrotational dust 
breaks down and it is necessary to give a prescription on how to proceed 
with them.
We followed the standard assumption, as suggested by the virial theorem,
that collapsing regions shrink to half of their maximal sizes;
somewhat unrealistically we pretended that such regions 
contract according to the irrotational dust evolution equations until that 
size is reached.
The collapsed regions themselves were then treated in two distinct ways:
firstly, by keeping 
them and letting all quantities retain the values that they had in the last
moment of collapse, and secondly by just removing them from the statistics.
%We believe that t
The second approach makes more sense since it 
is doubtful whether many of the observed photons would have passed through a 
collapsed region, 
and also because the strong anisotropies that can occur during collapse 
should not persist in the virialized regions;
nevertheless it is useful to have the other approach as well in order
to get an idea of how strongly our results depend on details of
modelling.
In order to check that our results do not come solely from collapsing
regions, we also performed computations in which we excluded any region
from the statistics as soon as it started to contract.
We will refer to these approaches as scenarios 1/2/3, respectively.

The starting point is a computation of the basic results of the averaging
process.
The time evolution of 
$\<\sqrt{\bar g}\>_\mathrm{mw}=\<\sqrt{(g_{11} + g_{22} + g_{33})/3}\>_\mathrm{mw}$
as computed according to Eq.~(\ref{g11}), does not differ substantially from
that of its EdS equivalent $t^{2/3}$ (the discrepancy is less than $15\%$
for the scenarios and time intervals that we consider here).
We present our further results mainly in the form of figures created by GNU
octave \cite{octave}.
In these figures we use a colour coding of blue/cyan/green for scenarios
1/2/3, respectively, with dashed lines for the quantities $H_\sharp$, $q_\sharp$
and $d_{S\sharp}$ and solid lines for the other quantities corresponding
to these scenarios; furthermore EdS values are indicated by red dash-dotted,
volume average results by solid yellow, perturbative results by dotted magenta
and $\L$CDM reference values by black dotted lines.

\del
The first three of them display 
and the remaining figures show quantities that are derived from these by
using formulas from Sec.~\ref{sedf}.

\begin{figure}[H]
\begin{center}
\includegraphics[width=16cm]{pp88allergbt.pdf}
\capt{12cm}{pp88allergbt}{Time evolution of 
                           $\<\sqrt{\bar g}\>_\mathrm{mw}/t^{2/3}$}
\end{center}
\end{figure}

\fref{pp88allergbt} displays the time evolution of 
$\<\sqrt{\bar g}\>_\mathrm{mw}=\<\sqrt{(g_{11} + g_{22} + g_{33})/3}\>_\mathrm{mw}$
as computed according to Eq.~(\ref{g11}), divided by the EdS value of $t^{2/3}$.
The blue line shows the result for the first scenario (collapsed
regions removed from the statistics) and the green (highest) line gives the 
same curve for the other scenario. 
While the quotient remains extremely close to 1 in the first case, even in 
the second case the deviation of the behaviour from the EdS values,
as indicated by the straight red line, is not large 
(note the scale of the vertical axis).
\enddel

\begin{figure}[H]
\begin{center}
\includegraphics[width=16cm]{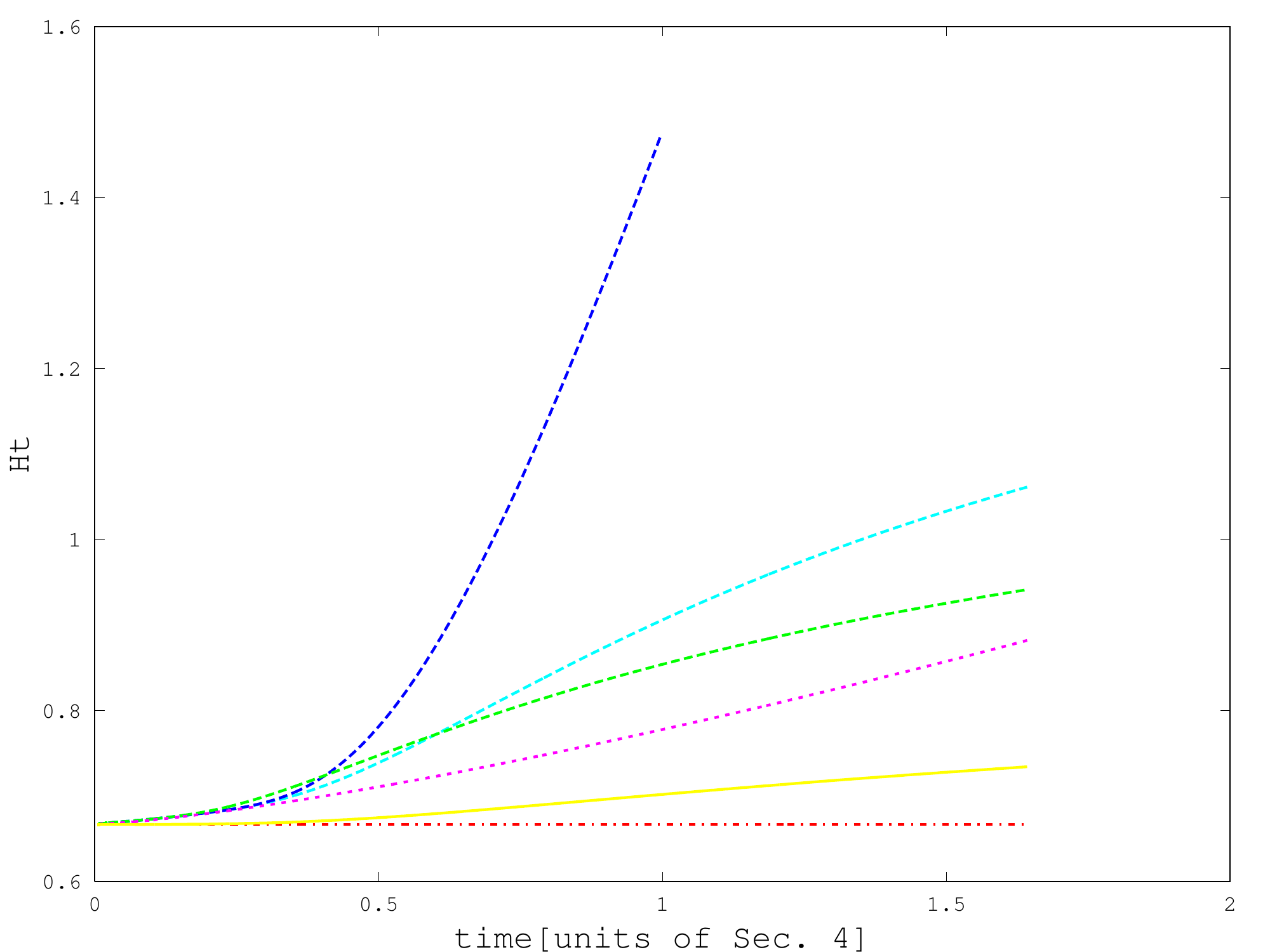}
\capt{12cm}{pp88allHtt}{Time evolution of $H t$}
%: Non-perturbative, perturbative and volume average}
\end{center}
\end{figure}

\fref{pp88allHtt} displays $Ht$ over the time $t$ for various versions
of the Hubble rate $H$.
The dashed lines in blue (highest), cyan (second) and green (third)
correspond to the results of the non-perturbative computations;
more precisely, they give $\<\Ht \>_\mathrm{pp} t$ as computed numerically
via Eq.~(\ref{Hnonpert}) for the scenarios 1, 2 and 3, respectively.
The fourth line (dotted, magenta) corresponds to the perturbative result 
(\ref{Hpert}), the fifth (solid yellow) line to $H t$ as computed 
%according to the 
via volume averaging, and the final red dash-dotted line shows the constant
EdS value of $H_\mathrm{EdS}t = 2/3$.

The strong deviations from the homogeneous case are a consequence mainly of
local anisotropy, by the following mechanism.
Consider a region $\cR$ characterized by some specific values of $\th$ 
and $\s_{ij}$ and 
pick a frame $\{\mathbf{e}_1, \mathbf{e}_2, \mathbf{e}_3\}$ in which
$\s_{ij}$ is diagonal.
Assume, without loss of generality, %for the sake of simplicity, 
that $\s_{11}> \s_{22}$ 
%has a positive eigenvalue along the $x$-direction and a negative eigenvalue 
%along the $y$-direction,
and that originally $\cR$ had the same diameters along the corresponding 
directions $\mathbf{e}_1$, $\mathbf{e}_2$.
Even though the overall volume expansion of $\cR$ is determined by $\th$,
it will expand faster along $\mathbf{e}_1$ and more slowly along $\mathbf{e}_2$,
%$y$-direction, 
%Since these directions do not change much over time, 
so that after a while $\cR$ will have a larger extension in the 
$\mathbf{e}_1$-direction than in the $\mathbf{e}_2$-direction.
A photon traversing $\cR$ along $\mathbf{e}_1$ will not only experience a 
stronger redshift \emph{per unit of time spent in} $\cR$ than one moving
along $\mathbf{e}_2$, but it will also spend more time in $\cR$.
The corresponding weighting that favors directions with stronger expansion
results in the effect that on average a photon traversing $\cR$ experiences 
a higher redshift than the volume expansion of $\cR$ would suggest.

\begin{figure}[H]
\begin{center}
\includegraphics[width=16cm]{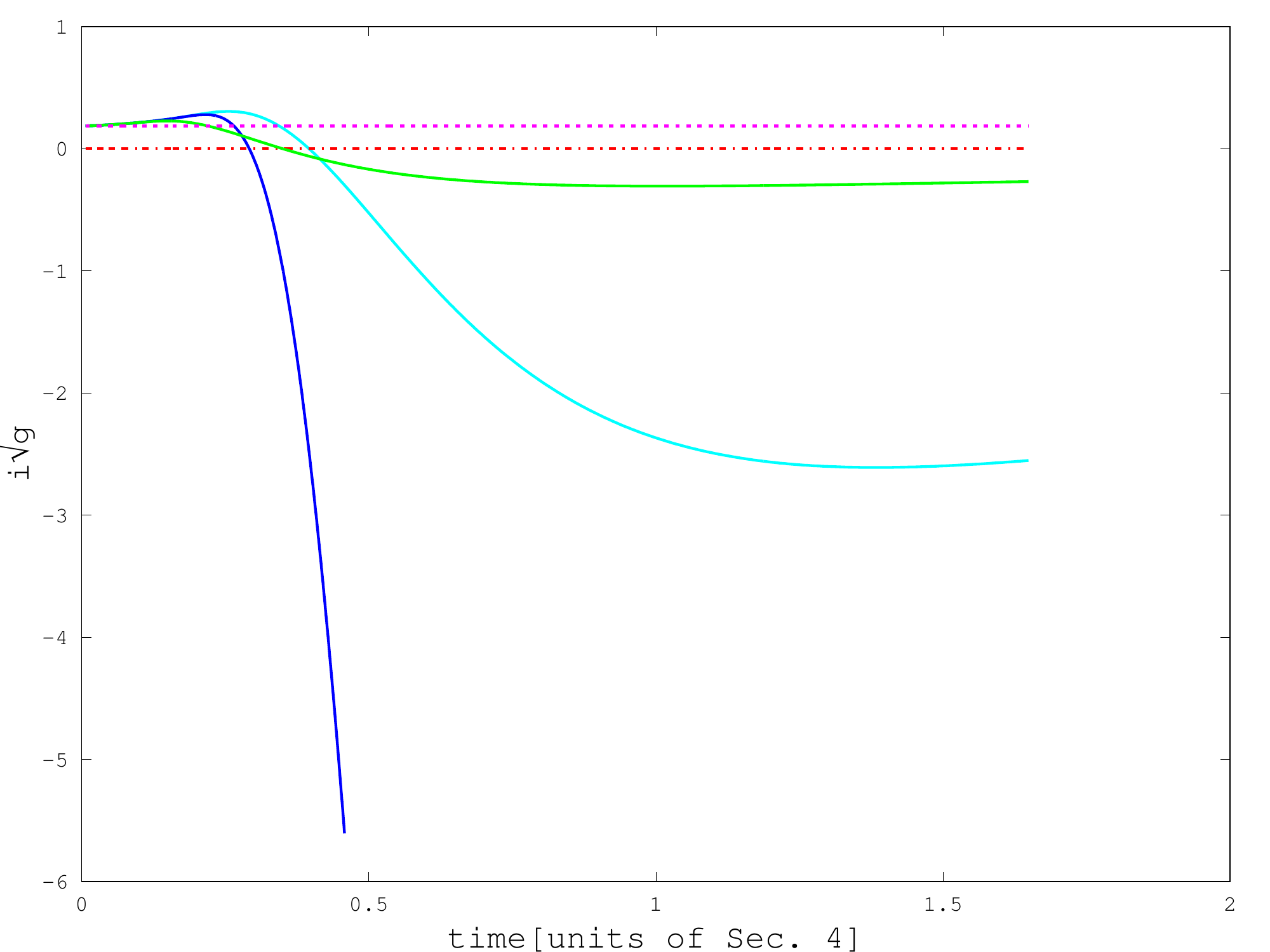}
\capt{12cm}{pp88alleirgb}{Time evolution of $i\sqrt{\bar g}$}
\end{center}
\end{figure}

In \fref{pp88alleirgb} the time evolution of $i\sqrt{\bar g}$
is displayed for our three non-perturbative scenarios; to be precise,
$\<i\>_\mathrm{pp}\<\sqrt{\bar g}\>_\mathrm{mw}$, i.e.~the mass-weighted
average of the right-hand side of Eq.~(\ref{ippnum}) 
is shown.
The sharply dropping blue line corresponds to the first scenario, the 
curved cyan line to the second one, and the mildly dropping green line
to the third one.
These results are contrasted with the perturbative result 
$i\sqrt{\bar g} \equiv 5/27$ and the EdS value of $i\sqrt{\bar g} \equiv 0$
as represented by the two horizontal lines (in dotted magenta and dash-dotted
red, respectively).
Here the differences between the perturbative and non-perturbative 
results are not only enormous in magnitude but also change the direction of 
the effect.
Once again the main contributions come from terms involving indicators of
local anisotropy such as $\s_{ij}$ and $r_{ij}$, as the form of the 
defining equation (\ref{iirrd}) suggests.

\begin{figure}[H]
\begin{center}
\includegraphics[width=16cm]{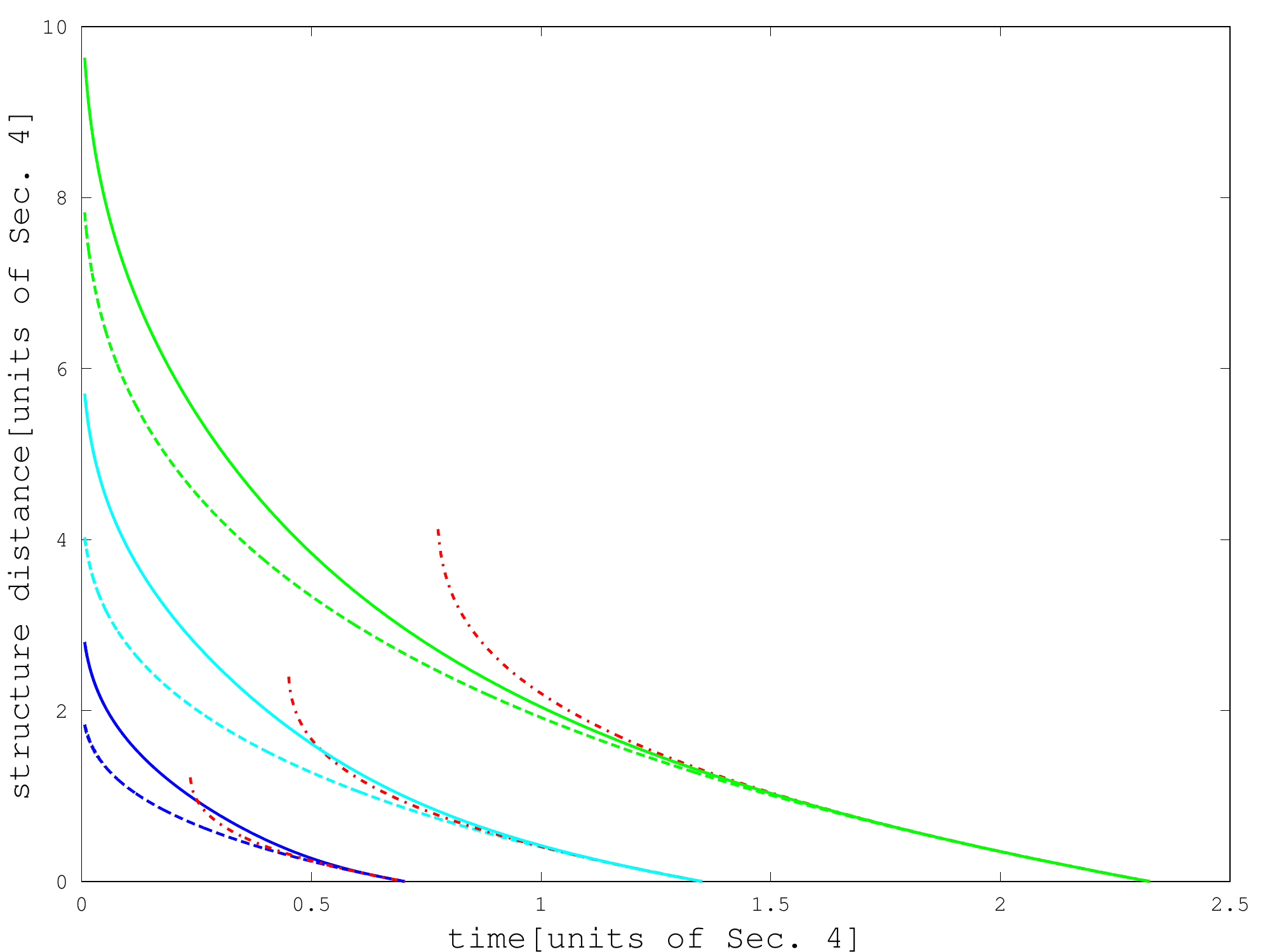}
\capt{12cm}{zd88alldt}{Structure distance over time}
\end{center}
\end{figure}

\fref{zd88alldt} differs from the previous ones by relying not only on
$t_e = t$ but also on $t_o$, the present age of the universe
expressed in the dimensionless units of Sec.~\ref{mwa}.
Here and elsewhere our choice was simply to take $t_o$ as the time at which
$\Ht t = 1$ (remember that $\Ht(t_o) = \Hd(t_o)$).
This is suggested by the fact that it seems to be a very good approximation
in the case of the $\L$CDM model and also close to lower bounds coming
from ages of globular clusters; in a more general analysis one should 
probably also allow for values of $H_ot_o$ somewhat above 1.
For our first scenario we find $t_o\approx 0.7$ in this way.
The three lines ending at that value show various versions of the structure
distance as functions of $t=t_e\in [0, t_o]$: 
the solid blue line shows $d_S$ itself, the dashed blue line below corresponds
to $d_{S\sharp}$, and the dash-dotted red line
that ``starts late'' corresponds to an EdS universe with the same value 
of $H_o$, which would have had a shorter lifetime up to now.
The other two triplets of lines correspond in an analogous way to the second
scenario, where $t_o\approx 1.35$, and to the third one with $t_o\approx 2.3$.

\begin{figure}[H]
\begin{center}
\includegraphics[width=16cm]{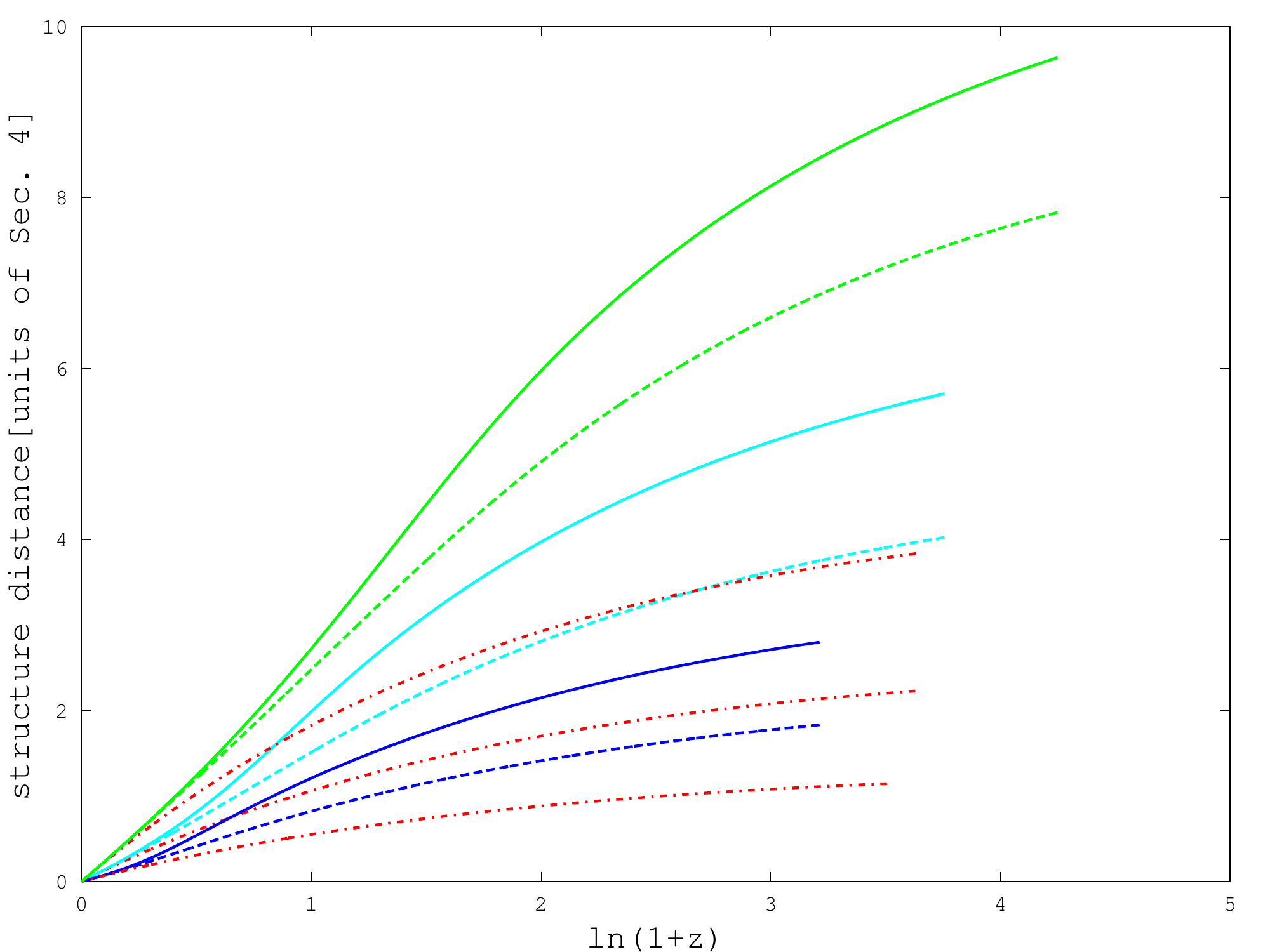}
\capt{12cm}{zd88alldz}{Structure distance over $\ln(1+z)$}
\end{center}
\end{figure}

For producing \fref{zd88alldz}, a plot of various versions of the 
structure distance over $\ln(1+z)$, the result of Eq.~(\ref{Hnonpert})
(as shown in \fref{pp88allHtt}) was integrated to get 
$\ln(1+z)$ as a function of $t$, and combined with the values
for the structure distance as displayed in \fref{zd88alldt}.
%the previous plot. Again t
Each scenario is represented by a triplet of lines starting with the same
slope which is lowest for the first and highest for the third scenario;
%The three lines that start with the lowest slope correspond to the first
%and the other three lines correspond to the second scenario, with
the colour and linestyle coding are the same as before.
This plot shows that $d_S^\mathrm{(EdS)} < d_{S\sharp} < d_S$, with 
differences of roughly the same size; i.e.~the effect of 
%passing from the EdS structure distance to $d_{S\sharp}$, i.e.~of 
a proper treatment of the second coefficient
$-[\ln(1+z)]\rdot$ in Eq.~(\ref{dSeq}) is of the same order of magnitude as
that of a proper treatment of the third coefficient, the quantity $i$.

\begin{figure}[H]
\begin{center}
\includegraphics[width=16cm]{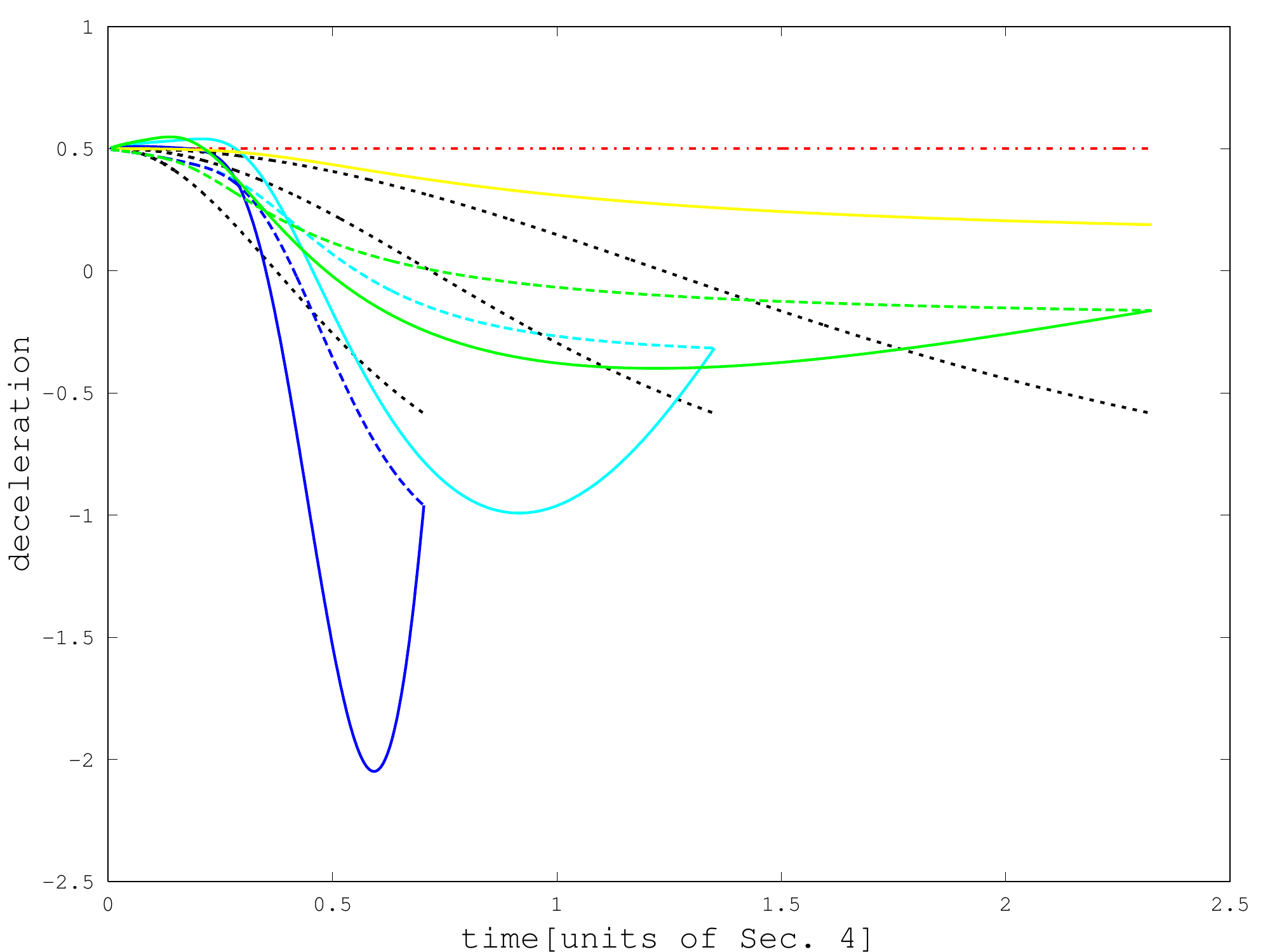}
\capt{12cm}{zd88allq}{Deceleration over time}
\end{center}
\end{figure}

\fref{zd88allq} displays various versions of the deceleration parameter 
over the time $t$.
The colour coding is the same as in the previous plots.
The dashed lines give $q_\sharp$ and the solid blue, cyan and green lines
represent $q_\mathrm{inf}$;
in each case the lines end at our choice for $t_o$.
The black dotted lines correspond to deceleration in the
standard $\L$CDM scenario with $\O_\L = 0.72$, with $t_o$ identified with the
present time.
Again the straight red dash-dotted line represents the EdS scenario, where 
$q\equiv 1/2$, and the yellow line which shows only a slight downward slope
displays the values that one gets via volume averaging.
Once again we see that the photon path prescription leads to strongly 
different results, with effects of roughly the same order coming from
the more precise treatments of the two non-trivial coefficients in
Eq.~(\ref{dSeq}).

While all the results presented so far refer to times and distances in terms
of the mathematically convenient but observationally meaningless units of 
Sec.~\ref{mwa}, the following plot uses standard units of years and parsecs.

\begin{figure}[H]
\begin{center}
\includegraphics[width=16cm]{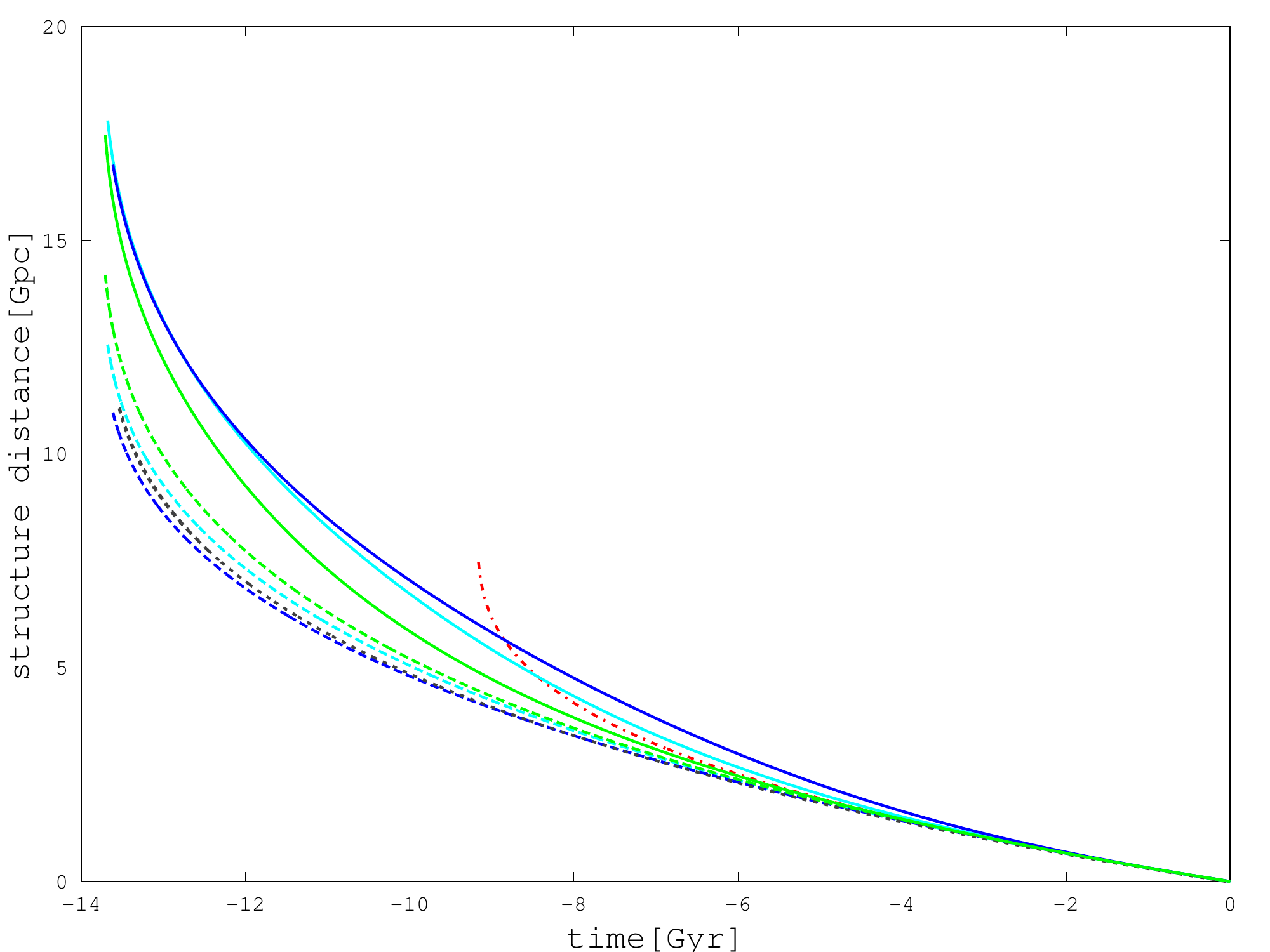}
\capt{12cm}{zd88alldtn}{Structure distance over time}
\end{center}
\end{figure}

\fref{zd88alldtn} is identical to \fref{zd88alldt} except for the 
normalization and the inclusion of a reference $\L$CDM curve (again as a
black dotted line).
% which was modified via $t\to t/t_o$, $d\to d/t_o$.
This figure shows that, with the correct scaling, the predictions of the
three different scenarios actually differ less than it appeared originally.
Somewhat surprisingly, $d_{S\sharp}$ is closer to the $\L$CDM values than
$d_S$ here; in particular our results for $d_S$ overestimate the distances
for early emission times.
We can make this discrepancy quite precise by computing the distance
to the last scattering surface from which the cosmic microwave background
stems (see also Ref.~\cite{Clarkson:2014pda}).
This is not completely straightforward because the stepwidth of our programs
is not fine enough for handling the time $t_\mathrm{ls}$ of last scattering
that corresponds to $z=1090$.
%taking $t_\mathrm{ls}$ as the time at which $z=1090$ and using our programs,
We have circumvented this obstacle by using a combination of our programs
and linear perturbation theory to find $t_\mathrm{ls}$,
noting that the solution of Eq.~(\ref{dSeq}) near $t=0$ takes the form
%of a Taylor series in $t^{1/3}$,
$d_S(t) = d_S(0) + d_S^{(1)}t^{1/3} + \co(t^{2/3})$,
checking that the numerical results for small $t$ are very well fitted by the
first two terms, and using them to get $d_S(t_\mathrm{ls})$.
Upon doing this and converting the result to standard units, we found
$d_S(t_\mathrm{ls}) \approx 20.7/20.9/19.8 ~\mathrm{Gpc}$ for scenarios 1/2/3,
respectively.
These numbers overestimate $d_S$ by almost 50\% compared to Planck results
\cite{Ade:2013zuv} of $13.9 ~\mathrm{Gpc}$
(see their Table 2 and use $d_S=r_*/\th_* [\mathrm{Mpc}]$), which is the
largest discrepancy from standard values that we found in the present work.
There are two possible explanations.
On the one hand, we have omitted the term $(1+z)^{-2}|\sopt|^2$
(related to Weyl focusing) in Eq.~(\ref{iirrd});
cf.~the discussion after Eq.~(\ref{Hnonpert}).
Inclusion of this term would make the shape of the function $d_S(z)$ flatter
and therefore more similar to the $\L$CDM reference curve.
On the other hand it is not clear whether the angular distance as inferred
from the Planck results really should be exactly the same one as that
computed via the Sachs equations.
The Planck results refer to finite physical distances at $t=t_\mathrm{ls}$,
whereas the Sachs equations refer to the intersection of the observer's
backward light cone with that timeslice.
In the homogeneous case this intersection will be perfectly spherical, but
in a realistic inhomogeneous universe it might be somewhat crumpled
(more like the surface of an orange), and the distance that corresponds
to a total length along that surface (which is what the Sachs equations
compute) will be somewhat larger.

\begin{figure}[H]
\begin{center}
\includegraphics[width=16cm]{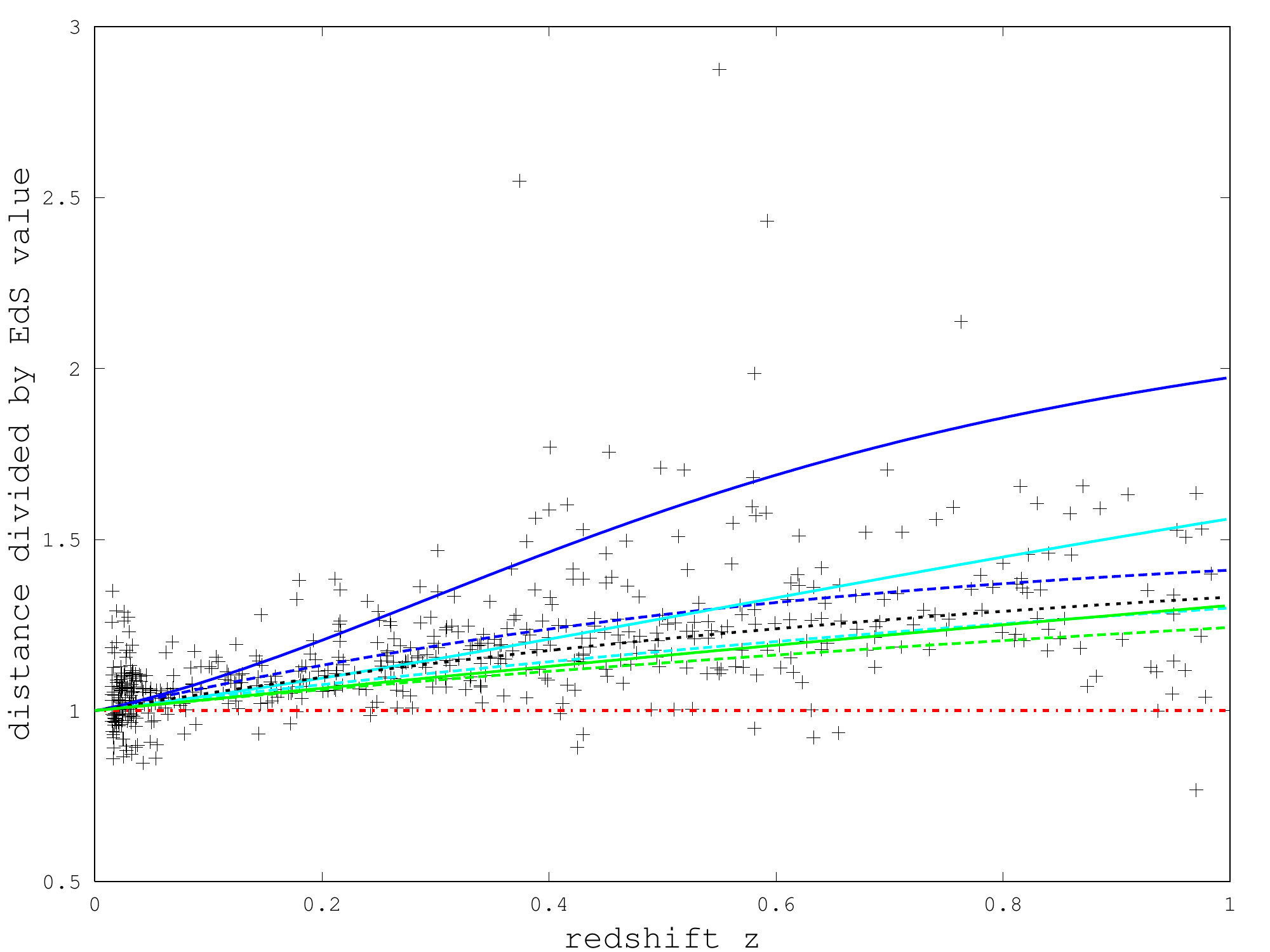}
\capt{12cm}{zd88alldznh}{Distance (normalized to EdS values) over $z$}
\end{center}
\end{figure}

\fref{zd88alldznh} displays, like %is a modified version of
\fref{zd88alldz}, distance over redshift, the changes being the normalization
of the distance to EdS values, the narrower range of $z$-values, the use of $z$ 
instead of $\ln(1+z)$, and the inclusion of the $\L$CDM scenario and supernova
data.
Again the red dash-dotted line corresponds to an EdS universe,
the black dotted one to a $\L$CDM universe with $\O_\L = 0.72$,
the solid lines to the observed structure distances $d_S$ for our three
scenarios, and the dashed lines to the values of $d_{S\sharp}$.
The black crosses mark the 551 supernovae from the Union2.1 compilation
\cite{Suzuki:2011hu} that have $z<1$, as taken from the Supernova Cosmology
project website \cite{Union21website}.
Both the second and the third scenario perform much better than the EdS case;
actually the $\L$CDM curve lies between the second and third scenario for
most of the redshift values shown in the plot, and the second one somewhat
overestimates the deviation from EdS.
The first scenario, in which collapsed regions are included
with the values for $[\ln(1+z)]\rdot$ and $i$ that they had in the last 
moment of collapse, overestimates these deviations even more strongly.
This suggests that our model would be improved by introducing
a smooth slowing of the collapse (as it happens in reality), with a 
corresponding smooth transition of $[\ln(1+z)]\rdot$ and $i$ to zero.
The fact that even our third scenario, in which we have suppressed the effects
from contracting regions, deviates strongly (and in the right direction) from
the EdS case demonstrates that such an improvement could not obliterate
the total effect of our treatment of inhomogeneities.

What have we seen up to now? 
Considering a universe with $\L = 0$ and with distributions 
of geometric quantities that follow directly from initial conditions based on 
a Gaussian distribution, and with photons that obey 
%distance formulas derived from 
the Sachs optical equations, we have shown that the following facts hold:
there is a time $t_o$ such that %$(Ht)_o = 1$;
an observer at that time sees a redshift-distance relation
remarkably similar to that predicted by the standard $\L$CDM scenario, and
if the observer analyses the data without taking into account the 
inhomogeneities, he will infer a Hubble rate $\Hd$ such that $\Hd t_o = 1$
and a deceleration parameter $q_\mathrm{inf}\approx -0.5$. % at that time.

We have already considered the time $t_\mathrm{ls}$ of last scattering
in our discussion of \fref{zd88alldtn}.
We can make a further, less ambiguous, statement on that era in the
following manner.
In our most realistic scenario (the second one), %in which 
%A rough estimate is
%$t_o \approx 1.35$ in the dimensionless units of Sec.~\ref{mwa},
%Let us now use this result to make a statement on the time $t_\mathrm{ls}$
%of last scattering, from which the cosmic microwave background stems.
%taking $t_\mathrm{ls}$ as the time at which $z=1090$ and using our programs,
$t_\mathrm{ls} \approx 5.3\times 10^{-5}$ in the %same units
dimensionless units of Sec.~\ref{mwa}
(with $t=0$ the instant at which the singularity would have occurred in a 
purely matter dominated universe).
At this time linear perturbation theory is still perfectly valid so that
we can compute density perturbations at last scattering with the help of
formulas (\ref{apert}) and (\ref{sevs}):
\beq
\({\D\r\0 \r}\)_\mathrm{ls} = \({\D (a^{-3})\0 a^{-3}}\)_\mathrm{ls}
= \2 t_\mathrm{ls}^{2\0 3}\, \D S 
= \2 \times (5.3\times 10^{-5})^{2\0 3}\times \sqrt{5}
\approx 1.6\times 10^{-3}.
\eeq
These are the density perturbations for the total matter, which are 
dominated by the ones for dark matter.
According to Eq.~(2.6.30) of Ref.~\cite{Weinberg:2008zzc}, the density
perturbations
of baryonic matter satisfy $\D\r_B/\r_B = 3\D T/T$, where $T$ is temperature;
using the commonly cited value of $10^{-5}$ for the relative temperature 
fluctuations in the CMB we find that the total density perturbations are 
roughly 50 times as large as those for the baryonic matter. 
This fits very well with the fact that dark matter decouples from photons
(hence clumps gravitionally) earlier than baryons. %this is a 
%The same feature is also required for structure formation, where 
Similar values for the ratios of the baryonic versus total density perturbations
are required for structure formation; see e.g.~Fig.~1 of 
Ref.~\cite{astro-ph/0503196}.
We can turn this argument around: from the density perturbations we see that
the time of last scattering cannot have occurred significantly before the
time $t_\mathrm{ls} \approx 5.3\times 10^{-5}$ that corresponds to 
$t_o \approx 1.35$.
But then it is clear that the inhomogeneities will have a significant 
impact on inferred Hubble and deceleration rates, so that the assumption that
a homogeneous universe (with or without a cosmological constant)
give correct predictions necessarily breaks down.
Conversely, since we do not require a non-zero $\L$ to account for present
observations the simplest assumption is to take $\L=0$.

%\newpage
\section{Discussion and outlook}
Let us start our discussion with a brief reiteration of our assumptions
and conclusions.
Considering a universe that 
\begin{itemize}
\item is matter dominated and obeys the Einstein equations,
\item in its early stages was very close to being spatially flat and 
homogeneous, with only Gaussian perturbations, and
\item has vanishing cosmological constant, $\L=0$,
\end{itemize}
we found that there is a time $t_o$ such that 
observervations made at that time and interpreted with formulas appropriate 
to the homogeneous case, would suggest
\begin{itemize}
\item an inferred Hubble rate $H_\mathrm{inf}$ such that 
   $H_\mathrm{inf}t_o \approx 1$,
\item an inferred deceleration parameter of $q_\mathrm{inf}\approx -0.5$, and
\item density perturbations at a redshift of 1090 that fit well with values
required at last scattering to lead to structure formation.
\end{itemize}
In other words, an observer at time $t_o$ in such a universe sees essentially 
what present day cosmologists see, even though $\L$ vanishes.
This is the consequence of a model that has only one parameter (the
overall scale) which can be adjusted.
Once this parameter has been fixed by any of the three quantities that
were just mentioned (and thus $t_o$ identified with the present age of the 
universe), the prediction for either of the other two provides
a highly nontrivial test. 
Our methods have performed very well on both of them.

In order to arrive at these results it is essential to consider the effects
of inhomogeneities on light propagation (not just on the evolution of
volumes), and to use a formalism that transcends perturbation theory.
The main steps involve the derivation of the differential equation 
(\ref{dSeq}) for the structure distance $d_S = (1+z) d_A$, and the computation
of the two non-trivial coefficients $\Ht = -[\ln(1+z)]\rdot$ and $i$ that
occur in this equation.
In the spatially flat homogeneous case $\Ht$ is just the usual Hubble rate 
and $i=0$;
otherwise each of these coefficients contributes significantly, 
with effects of roughly the same magnitude, to the deviations in the values
of $d_S$, $H_\mathrm{inf}$ and $q_\mathrm{inf}$.
The main source of discrepancies from FLRW universes is the local
anisotropy, as encoded in the dust shear $\s_{ij}$ and the traceless part
$r_{ij}$ of the Ricci tensor, and not so much the inhomogeneity which manifests
itself by variations of the expansion rate $\th$ and the spatial Ricci scalar
$R$.
While Eq.~(\ref{dSeq}) is valid in an arbitrary geometry in which 
photons follow light-like geodesics, the subsequent computations
required a number of approximations:
\begin{itemize}
\item The matter was modeled as irrotational dust.
While this is an excellent approximation during expansion, it would not 
permit stable structures such as galaxies and clusters as the results of
collapse. 
Our way of treating this problem, by simply assuming that collapse holds at
half the maximum size (or ignoring collapsing regions altogether),
is certainly somewhat ambiguous. 
In particular, the differences between the three variants that we chose show 
that the results do depend on such details;
at the same time our third scenario demonstrates that deviations from
the homogeneous case do not stem exclusively from collapse.
As we argued in the discussion of \fref{zd88alldznh}, a smoother
transition to the virialized state in our framework would probably lead to
even better agreement with observations.
\item We have replaced statistical quantities by their
expectation values in order to arrive at a description in which distance
can be seen as a function of redshift, as in homogeneous models
(cf.~the first paragraph of Sec.~\ref{secppav}).
From the set of supernova data it is clear that this is a gross 
oversimplification.
\item We assumed a distribution of photon paths in $\mathbf{x}$--space (the 
space in which our matter is at rest, which starts out as being almost 
perfectly euclidean) that was the same as if the photons moved along straight 
lines in that space.
\item While exact evolution equations were used for the local scale factor $a$,
the shear $\s_{ij}$ and the Ricci scalar $R$, the evolution of the traceless 
part $r_{ij}$ of the Ricci tensor was simplified by ignoring the right-hand 
side of Eq.~(\ref{rhatev}). 
\item In our analysis of expressions that arise upon taking photon path 
averages, we have neglected terms of quadratic or higher order in $\g_{ij}$ 
(a scaled version of the traceless part of the metric $g_{ij}$).
\item For reasons that we discussed after Eq.~(\ref{Hnonpert}) we %completely 
ignored Weyl focusing, i.e.~the contribution of the optical shear 
$\s_\mathrm{opt}$.
\item For the numerical treatment the time axis and the probability 
distribution for the background parameters were discretized. 
The resulting errors are, however, much smaller than those coming 
from the other approximations.
\end{itemize}
%While these items look quite reasonable,
Unfortunately the second and third item %unot so
are not as harmless as they originally seemed.
%Our computations in Sec.~\ref{pertres} respect the essential terms at
%second order perturbation theory, but not all of the approximations do.
%In particular, u
Upon replacing $i$ and $\Ht$ by their expectation values, we
have introduced errors $\D i$ and $\D\Ht$ which have vanishing expectation
values and are of first order.
These lead to errors $\D d$ and $\D\ln(1+z)$ of the same type.
The transition from $d(t)$ and $\ln(1+z)(t)$ to $d(z)$ then generates products
of errors which are of second order and nonvanishing expectation value.
The approximation introduced in the second paragraph of
Sec.~\ref{secppav} probably leads to similar problems,
%violates second order perturbation theory,
whereas our computations in Sec.~\ref{pertres} respect the essential terms at
second order perturbation theory.

A general $n^\mathrm{th}$ order term is an $n$-fold product of $C$ (or,
equivalently, the Newtonian potential $\Phi$) or its derivatives,
in such a way that typically the $n^\mathrm{th}$ order term has a total of
up to $2(n-1)$ spatial derivatives more that the first order term
(see Ref.~\cite{Clarkson:2011uk} for a detailed discussion).
While $C$ itself is small, $\6^2 C$ can be large;
for example, density perturbations are of this type.
In particular, among the terms contributing to the redshift-distance relation
at second order, the largest ones that we find are proportional to
$\<(\6^2 C)^2\>$.
\del
., but according to
Refs.~\cite{1207.1286,1207.2109} terms of this type precisely cancel each
other, leaving only much smaller contributions from subdominant terms.
\enddel
However, according to the two independent groups that have performed complete
computations up to second order
\cite{1207.2109,Umeh:2014ana,1207.1286,BenDayan:2013gc}, terms of that type
%$\<(\6^2\Phi)^2\>$, which account for the second order results in the present work,
cancel out completely and subleading terms give corrections of an
order of magnitude of only around $10^{-4}$.

This can be explained in the following way.
In our approach, using the synchronous gauge, the whole setup relies
on expressing quantities in terms of the entries (or eigenvalues) of the
matrix $S_{ij} = \6_i\6_j C$;
to be precise, the $n^\mathrm{th}$ order contribution to any of the quantities
$a$, $\hat \s$, $\hat R$ and $\hat r$ is homogeneous of degree $n$ in $S$.
Terms of this type also produce the dominant contribution to the deviation of
the metric from the FLRW case.
But terms of (schematically) type $\6^{2n} C^n$ in $g_{ij}$ give rise to terms
of type $\6^{2n+2} C^n$ in the curvature, which must all cancel.
This means that the part of the spatial metric consisting of the highest
derivatives is flat, implying that it is possible to reparameterize the
spatial slices in such a way that the metric no longer contains the
$\6^{2n} C^n$ terms.
Hence any approximation in the synchronous gauge that does not respect the
precise structure of the $\6^{2n} C^n$ terms introduces errors that are
potentially larger than the physical effects from the inhomogeneities.
Since our approach suffers from this problem, it does not provide a conclusive
argument that the standard $\L$CDM picture require modification.
Nevertheless it is intriguing how well it appears to perform -- after all,
one would expect mere %the consequences merely of
errors to result in random %consist of arbitrary
nonsense rather than something that closely resembles observations.
Besides, standard perturbation theory cannot be trusted either:
higher order terms are not smaller than first order terms
\cite{Clarkson:2011uk}, and the real universe features shell crossings and
vorticity, which do not occur in a purely perturbative modelling of an
irrotational dust universe, but whose effects are taken into account by
the approach to virialization in the present framework.

\del
Thus it seems plausible that the cancellations of the $\6^{2n} C^n$ terms
in the redshift--distance relation occur to all orders in perturbation theory,
indicating that our approximations may introduce terms of a type that simply
do not occur in the full expansion.
\enddel

\noindent
{\it Acknowledgements:} 
It is a pleasure to thank Anton Rebhan and Dominik Schwarz for
helpful discussions, and Phil Bull for email correspondence.
While a lengthy refereeing process led to some clarifications, the arguments
in the last few paragraphs that shed doubt on the approach were developed
by myself during a stay at Bielefeld university, for whose hospitality I am
very grateful.

\newpage

%\bye

%\small

%\enddel

\bye